\documentclass[a4paper,fleqn]{cas-sc}

\usepackage{subcaption}
\usepackage{hyperref}
\hypersetup{
    colorlinks=true,     
    linkcolor=blue,      
    citecolor=blue,      
    urlcolor=blue        
}

\usepackage[numbers,sort&compress]{natbib}

\def\tsc#1{\csdef{#1}{\textsc{\lowercase{#1}}\xspace}}
\tsc{WGM}
\tsc{QE}
\tsc{EP}
\tsc{PMS}
\tsc{BEC}
\tsc{DE}

\ExplSyntaxOn
\cs_gset:Npn \__first_footerline:
  { \group_begin: \small \sffamily \__short_authors: \group_end: }
\ExplSyntaxOff 

\begin{document}
\let\WriteBookmarks\relax
\def\floatpagepagefraction{1}
\def\textpagefraction{.001}
\shorttitle{Nonlinear wave scattering modeling}
\shortauthors{Qianyue Chen et~al.}

\title [mode = title]{On Modelling the Scattering of Highly Nonlinear Transient Water Waves Due to Surface-Piercing Cylinder. Part I: Formulations and Verifications}                   

\author[1,2]{Qianyue Chen}[style=chinese]
\ead{qianyue.chen@connect.polyu.hk}
\credit{Methodology, Software, Validation, Formal analysis, Writing}

\author[1,2]{Jinghua Wang}[style=chinese,orcid=0000-0003-1821-2362]
\ead{jinghua.wang@polyu.edu.hk}
\credit{Conceptualization, Methodology, Supervision, Funding acquisition, Project administration}

\author[1]{Daimeng Shang}[style=chinese,orcid=0009-0000-4817-3566]
\cormark[1]
\ead{daimeng8.shang@connect.polyu.hk}
\credit{Data curation, Software, Validation, Writing}

\author[1]{Xianlong Lin}[style=chinese,orcid=0000-0002-9550-5794]
\ead{xianlong.lin@connect.polyu.hk}
\credit{Investigation, Resources, Validation}

\affiliation[1]{organization={State Key Laboratory of Climate Resilience for Coastal Cities, Department of Civil and Environmental Engineering, The Hong Kong Polytechnic University},city={Hong Kong SAR},country={China}}

\affiliation[2]{organization={Shenzhen Research Institute, The Hong Kong Polytechnic University},city={Shenzhen},country={China}}

\cortext[cor1]{Corresponding author}

\begin{abstract}
The presence of extreme ocean waves poses substantial threats to the safety of marine engineering applications and navigational activities.  Accurate prediction of the nonlinear wave loads on offshore structures in large-diameter cylindrical configuration, such as monopile wind turbines and platform columns, requires a robust solution to the high-order wave diffraction problem. This paper introduces a novel, highly nonlinear time-domain numerical pseudo-spectrum model based on potential flow theory. By leveraging the Fredholm integral equation of the second kind and Legendre polynomials, a distinct efficient solver for computing the velocity potential on the cylinder surface and vertical velocity at free surface is developed. The present method is verified by comparing with a direct approach based on Fourier transform and established theoretical models for simulating the scattering of small steepness uniform waves. The convergence and computational efficiency of the present method is also demonstrated. The validation of the new model for simulating highly nonlinear transient water waves will be reported in Part II. The proposed model provides a powerful and efficient new framework for analyzing high-order nonlinear wave diffraction and hydrodynamic loads, with significant implications for both scientific research and the engineering design of coastal and offshore structures.
\end{abstract}




\maketitle

\section{Introduction}\label{Introduction}

The interaction of surface water waves with a vertical, surface-piercing circular cylinder has been a cornerstone of ocean engineering research for decades. The dominant hydrodynamic regime is governed by the dimensionless ratio $D/\lambda$, where $D$ is the cylinder diameter and $\lambda$ is the incident wavelength \cite{Chakrabarti1987,OtsukaIkeda1996,Linton2001}. When $D/\lambda<0.2$, the cylinder is slender relative to the wave scale and the flow lies in the drag-inertia regime \cite{Chakrabarti1987,OtsukaIkeda1996,Linton2001}. Depending on the Keulegan-Carpenter number $KC=UT/D$, with $U$ the wave-induced velocity amplitude and $T$ the wave period, see \cite{KeuleganCarpenter1958}, the forcing is dominated by viscous drag ($KC>20$), inertia effects ($KC<5$), or a combination of both ($5<KC<20$) \cite{SwanSheikh2015}. In this regime, the cylinder produces negligible disturbance to the incident wave field, apart from local flow separation and wake formation, and the total force is routinely estimated using Morison's equation \cite{Morison1950}, which employs undisturbed water-particle kinematics.

However, nonlinear wave phenomena, such as wave--wave interactions, wave--seabed interactions, and breaking, introduce additional force components that Morison's equation, which is often used in conjunction with first-order wave theory \cite{Shank1970}, cannot capture. The associated higher-order force oscillations lead to phenomena commonly referred to as, for example, `springing' \cite[see e.g.,][]{VidicPerunovicJensen2005,ShaoFaltinsen2014,PanTi2025a,PanTi2025b}, `ringing' and secondary load cycles (SLCs) \cite[see e.g.,][]{GrueBjorsholStrand1993,FaltinsenNewmanVinje1995,ChaplinRaineyYemm1997,GrueHuseby2002,FitzgeraldTaylorTaylorGriceZang2014,PaulsenBredmoseBinghamJacobsen2014,RiiseGrueJensenJohannessen2018,GhadirianBredmose2019,LiZhangLiuFanZang2022,HlopheAdcockDingZangDaiTangTaylor2025}, as well as impulsive slamming \cite[see e.g.,][]{vonKarman1929,Wagner1932,Goda1966,WienkeOumeraci2005,PaulsenSonnevilleMeulenJacobsen2019,GhadirianPierellaBredmose2023}. All of these effects can significantly accelerate structural fatigue and thereby undermine structural integrity. The higher-order force harmonics in springing, ringing, and SLCs originate from two distinct sources: the bound components already present in the incident nonlinear wave and the additional nonlinear interactions between the wave and the cylinder itself. Several higher-order force models have been developed to account for these contributions, including second-order force models \cite{Newman1996}, the FNV models \cite{FaltinsenNewmanVinje1995,KristiansenFaltinsen2017,TaylorTangAdcockZang2024}, and Stokes-type harmonic fitting models \cite{ChenDuffour2018,HlopheAdcockDingZangDaiTangTaylor2025}. Slam loads arise specifically when breaking waves impact the cylinder, producing transient, high-magnitude forces that differ markedly from conventional hydrodynamic loading. In recent years, Computational Fluid Dynamics (CFD) models have become an important tool for investigating these transient breaking-wave effects and have been shown to reproduce not only impulsive forces but also the associated flow separation and wake characteristics \cite{ChenZang2014,GhadirianBredmoseDixen2016,KamathChellaBihsArntsen2016,AggarwalTomaselliChristensenBihs2020,ChenMaDongMaTaiNiu2020,AliyarDucrozetBouscasseVenkatachalamFerrant2022,YangStansbyRogersBuldakovStagonasDraycott2023,GaoZhangZhang2024,YangXuTanZhangWangTkalich2026}. In some cases, hydroelastic effects in flexible cylinders have also been considered. These account for the radiated waves generated by structural vibrations and are important for a comprehensive analysis of structural dynamics, including added mass and hydrodynamic damping, as well as vortex-induced vibrations (VIV) and the associated lock-in phenomena \cite{ChenZangTaylorSunMorganGriceOrszaghovaRuiz2018,XuWang2021,JiangelMoctarZhang2024,PanTi2025a,PanTi2025b,ZhangelMoctarJiang2026}.

By contrast, when the cylinder is large ($D/\lambda>0.2$), the flow enters the diffraction regime and a significant disturbance of the incident wave field is generated \cite{Chakrabarti1987,OtsukaIkeda1996,Linton2001}. This disturbance manifests as a radially propagating scattered wave field, and both the applied hydrodynamic forces and the maximum crest elevation must therefore incorporate the contributions of the scattered waves. Linear diffraction theory, with analytical solutions for infinite and finite water depths provided by \cite{Havelock1940} and \cite{MacCamyFuchs1954}, respectively, remains widely employed in engineering applications due to its simplicity \cite[see e.g.,][]{ChildVenugopal2010,ChatjigeorgiouKorobkinCooker2016,RoyChakrabortyMandal2017,LiLiu2019,KangZengCuiChen2024}. However, this approach rests on the assumption of linear incident waves and therefore provides only a simplified representation of the actual diffraction process. In reality, ocean waves are often nonlinear, and the resulting diffraction can be considerably more complex than that predicted by linear theory. A series of laboratory experiments has demonstrated that linear diffraction theory can substantially under-predict the amplification, relative to the undisturbed incident wave, of both crest elevations and wave loads \cite{Kriebel1992,NiedzweckiDuggal1992,StansbergNielsen2001,StansbergBraaten2002,Nielsen2003,KristiansenBaarholmStansberg2004,StansbergBaarholmFokkGudmestadHaver2004,StansbergKristiansen2005}. In particular, \cite{StansbergKristiansen2005} reported underestimations of crest heights by as much as $75\%$ for steep incident waves. These experimental studies further identified the nonlinear components of the incident wave field, their interactions with the cylinder, and mutual interactions among incident and scattered wave components as important mechanisms governing the resulting wave elevation and hydrodynamic loading \cite[see e.g.,][]{SwanSheikh2015}.

To overcome the limitations of linear diffraction theory, a variety of higher-order and fully nonlinear numerical models have been developed. Frequency-domain approaches include second-order models \cite{KimYue1989,Kriebel1990,ChauEatockTaylor1992,TengKato1999} and third-order models \cite{MalenicaMolin1995,MolinKimmounRemyChatjigeorgiou2014,TengKato2002}. Time-domain approaches include finite element models \cite{MaWuEatockTaylor2001a,MaWuEatockTaylor2001b}, boundary element methods \cite{IsaacsonCheung1992,Ferrant1998,BaiEatockTaylor2007,BaiTeng2013,ZhouNingTengBai2013,DuanChenZhao2015}, spectral element methods \cite{EngsigKarupMonteserinEskilsson2017,MonteserinEngsigKarupEskilsson2018}, Green--Naghdi models \cite{HayatdavoodiNeillErtekin2018,HayatdavoodiErtekin2023,SuTengGou2024,SuGouZhouTengLi2026}, and weak-scatterer solutions \cite{TongShaoBingham2026}. Among these approaches, boundary element methods (BEMs) have played an important role in nonlinear free-surface and wave--structure interaction problems because the governing Laplace equation can be reduced to the boundaries of the fluid domain. Early work by \cite{GrilliSkourupSvendsen1989} demonstrated that a higher-order BEM coupled with high-order time integration could efficiently simulate highly nonlinear water-wave propagation, run-up, breaking, and transient wave generation. When integrated with Lagrangian or Mixed Eulerian--Lagrangian (MEL) free-surface tracking, BEM formulations provide a powerful framework for resolving fully nonlinear free-surface evolution and wave--body interactions \cite{MeiStiassnieYue2005,MeiStiassnieYue2018,XueXuLiuYue2001,LiuXueYue2001}. Subsequent developments introduced implicit time-domain formulations to improve numerical stability and accuracy in nonlinear radiation and wave--body interaction problems \cite{SungChoi2010,GuerberBenoitGrilliBuvat2012}. Fully nonlinear interactions between transient waves and three-dimensional vertical cylinders have also been investigated using higher-order BEM formulations \cite{ZhouWuMeng2016}. More recently, Arbitrary Lagrangian--Eulerian regularized boundary-integral formulations have been developed to alleviate the mesh distortion associated with strongly nonlinear free-surface deformation while retaining the advantages of boundary-integral methods \cite{TsaoKees2023}.

The importance of nonlinear diffraction has been demonstrated by both higher-order and fully nonlinear simulations. Second-order contributions can substantially amplify wave run-up, pressure, and hydrodynamic forces compared with linear predictions, particularly for moderate-to-steep incident waves \cite{ChauEatockTaylor1992,LiuXueYue2001}. Unlike first-order pressure components, higher-order pressure contributions may exhibit substantially different depth dependence because of the coexistence of bound and freely propagating modes. At stronger nonlinearities, fully nonlinear effects become increasingly important. \cite{Ferrant1998} simulated the diffraction of finite-amplitude regular waves by a vertical cylinder and demonstrated substantial nonlinear amplification of the local free-surface elevation. Such behavior is associated with nonlinear free-surface interactions that generate higher harmonics and free super-harmonic waves, whose subsequent diffraction and interaction around the cylinder cannot be fully represented by linear or low-order theories. Fully nonlinear BEM studies have further extended these analyses to transient wave--structure interactions. For example, \cite{ZhouWuMeng2016} investigated the interaction of a fully nonlinear solitary wave with a three-dimensional vertical cylinder, demonstrating the capability of higher-order BEM to resolve transient run-up, forces, and body responses. For focused wave groups, \cite{BaiEatockTaylor2007} demonstrated that extreme run-up and hydrodynamic loading depend strongly on the prescribed focusing position because of nonlinear interactions between the incident and scattered wave fields. These findings collectively underscore the necessity of nonlinear numerical models for predicting extreme wave loading. Nevertheless, existing fully nonlinear investigations have predominantly considered unidirectional regular waves, solitary waves, or long-crested focused wave groups. The nonlinear diffraction of three-dimensional, directionally focused transient waves by a large surface-piercing vertical cylinder remains much less explored.

Over the past few years, with the rapid global expansion of monopile offshore wind energy \cite{KoivistoGeaBermudezSorensen2020}, accurate prediction of transient and highly nonlinear extreme waves has become increasingly important for offshore engineering applications. Pseudo-spectral methods have emerged as powerful and efficient tools for simulating nonlinear water waves, particularly for investigating their evolution over large spatial and temporal scales \cite{DommermuthYue1987,WestBruecknerJandaMilderMilton1987}. This class of models is particularly well suited to generating nonlinear free-propagating waves in realistic three-dimensional seas \cite{WangMaYanLiang2021}. Previous studies have revealed distinct dynamic characteristics of extreme wave events, including bandwidth broadening associated with transverse modulation \cite{TrulsenDysthe1997}, directional focusing waves reaching substantially greater steepness before breaking than their unidirectional counterparts \cite{JohannessenSwan2001,JohannessenSwan2003,SwanJohannessenBateman2001}, and an extended lifespan of extreme events \cite{HeWangHeLiFengChabchoub2025}.

When these highly nonlinear, short-crested extreme waves interact with structures, the underlying physics can differ significantly from idealized long-crested scenarios. In directional seas, wave energy converges from multiple directions, altering the spatial distribution and nonlinear interaction of incident and scattered wave components around the cylinder. This can lead to asymmetric run-up distributions and distinct high-frequency hydrodynamic loading compared with unidirectional waves \cite{XiaoLiuWuYue2013,LiWangLiu2014a,WangMaYan2021,WangMaYangGaoWu2022}. The importance of such effects has been highlighted through laboratory experiments \cite{LiWangLiu2012,LiWangLiu2014a} and CFD simulations \cite{BihsChellaKamathArntsen2017,AliyarDucrozetBouscasseVenkatachalamFerrant2022,KangZhangYanJuZhuGu2024,SalisHuLuoRealiManenti2024}. Nevertheless, incorporating a vertically deployed surface-piercing cylinder into conventional pseudo-spectral formulations remains challenging because these methods are naturally formulated for open free-surface domains and do not directly accommodate internal vertical boundaries \cite{Grue2008,ZhuangWan2020}. Existing fully nonlinear boundary-integral approaches can accurately resolve wave--body interactions \cite{ZhouWuMeng2016,TsaoKees2023}, but their repeated boundary discretization and solution of the resulting boundary-value problem can become computationally demanding for large-scale three-dimensional transient-wave simulations. Therefore, a comprehensively verified numerical framework that retains the efficiency of pseudo-spectral wave models while accurately resolving nonlinear diffraction by a vertical cylinder is highly desirable. Bridging this methodological gap constitutes the primary motivation for the present study.

This paper aims to address this gap by presenting a novel, fully nonlinear time-domain model based on a well-established pseudo-spectral framework \cite{Wang2025}. The principal objective is to develop an efficient formulation for incorporating the boundary-value problem associated with a vertical surface-piercing cylinder into the pseudo-spectral framework, thereby enabling nonlinear wave diffraction to be resolved without sacrificing its computational advantages for large-scale wave simulations. Sections \ref{sec2} and \ref{sec3} detail the mathematical formulation and numerical implementation of the proposed method. Section \ref{sec4} examines the spatial discretization and convergence characteristics of the numerical scheme. Section \ref{sec5} provides rigorous verification against analytical diffraction solutions, followed by concluding remarks in Section \ref{sec6}. Building upon the verified numerical framework developed in the present Part I, Part II will provide experimental validation and extend the model to three-dimensional transient wave diffraction problems over broader hydrodynamic regimes.

\section{Mathematical formulation}\label{sec2}

\subsection{Coordinate system, governing equation and boundary conditions}\label{sec2.1}
\begin{figure}
	\centering
	\includegraphics[width=1\textwidth]{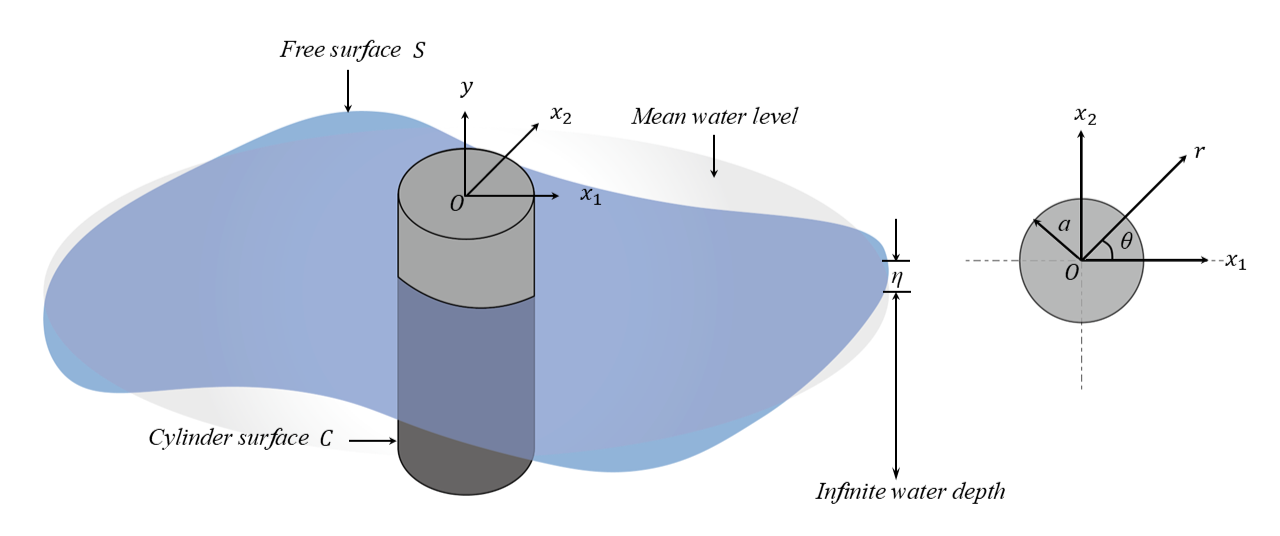}
	\caption{The sketch of coordinate system and computation domain.}
	\label{fig1}
\end{figure}

The fully nonlinear wave-cylinder interaction problem defining the fluid motion is considered. As shown in figure \ref{fig1}, the three-dimensional problem is defined in two coordinate systems: the right-handed Cartesian coordinates $O_{x_1 x_2 y}$, in which $\boldsymbol{x}=\left(x_1, x_2\right)$ are measured horizontally, and the polar coordinates $O_{\rho \theta y}$. The $y$ axis points vertically upward and $y=0$ denotes the mean water level. The radial axis $\rho$ overlaps the $x_1$ axis when $\theta=0^{\circ}$.

The fixed cylinder is located in the center of the domain ($\boldsymbol{x}=0$ or $\rho=0$) with its radius $a$; and the surface of the cylinder is denoted by $C$. Let $t$ denote the time and $y=\eta(\boldsymbol{x}, t)$ be the elevation of the free surface $S$. Thus, we give the water body $\Omega=\{(\rho, \theta, y) \mid \rho \geq a, y \leq \eta\}$.
The intersection of $S$ and $C$ is denoted as $L$. Unless otherwise specified, the variables used are dimensionless, i.e., the distance $(\boldsymbol{x},y,\rho\ \text{and}\ \eta )$ is multiplied by the peak wave number $k_p$, and the time $t$ by the circular frequency $\omega_p=\sqrt{gk_p}$, where $g$ is the gravitational acceleration.

The flow is assumed to be incompressible and irrotational, and potential theory is used to describe the flow with the velocity potential $\Phi(\boldsymbol{x},y)$ being introduced. The governing equation for the water body is in the form of Laplace function: 

\begin{align}\label{eq2.1}
\begin{split}
\nabla^2 \Phi+\frac{\partial^2 \Phi}{\partial y^2}=0\ \text{in}\  \Omega,
\end{split}
\end{align}

\noindent where $\nabla=\left(\partial _{x_1}, \partial _{x_2}\right)$ denotes the horizontal gradient operator and $\Phi$ is rescaled by $\sqrt{k_p^3/g}$.

The boundary conditions of the free surface are the following:

\begin{align}\label{eq2.2}
\begin{split}
\frac{\partial \eta}{\partial t} + \nabla \Phi\cdot \nabla \eta-\frac{\partial \Phi}{\partial y}=0\ \text{on}\  S,
\end{split}
\end{align}

\begin{align}\label{eq2.3}
\begin{split}
\frac{\partial \Phi}{\partial t}+\frac{1}{2}\left[|\nabla \Phi|^2+\left(\frac{\partial \Phi}{\partial y}\right)^2\right]+\eta=0\ \text{on}\  S.
\end{split}
\end{align}

Since the cylinder is assumed to be fixed and non-moving, the velocity on the cylinder surface is zero:

\begin{align}\label{eq2.4}
\begin{split}
\frac{\partial \Phi}{\partial n}=0\ \text{on}\  C.
\end{split}
\end{align}

\noindent where $\boldsymbol{n}$ is the unit normal vector that points outward of the fluid body.

\subsection{The prognostic equations}

Let $\phi(\boldsymbol{x}, t)=\Phi(\boldsymbol{x}, y=\eta(\boldsymbol{x}, t), t)$ be the velocity potential at the free surface, and $\varphi(\theta,y, t)=\Phi(\rho=a, \theta,y, t)$ be the velocity potential at the cylinder surface. 

After introducing the free surface vertical velocity $V=\partial_{n}\Phi \sqrt{1+|\nabla \eta|^2}$, the kinematic and dynamic boundary conditions of Eqs. (\ref{eq2.2}) $\&$ (\ref{eq2.3}) can be reformulated as:

\begin{align}\label{eq2.5}
\begin{split}
\frac{\partial \eta}{\partial t}=V,
\end{split}
\end{align}

\begin{align}\label{eq2.6}
\begin{split}
\frac{\partial \phi}{\partial t}  + \eta+ \frac{1}{2} \left( \left| \nabla\phi \right|^{2} - \frac{\left( V + \nabla\eta \cdot \nabla\phi \right)^{2}}{1 + \left| \nabla\eta \right|^{2}} \right) =0.
\end{split}
\end{align}

The free surface elevation and free surface velocity potential can be updated through Eq. (\ref{eq2.5}) and Eq. (\ref{eq2.6}) when $V$ is known, which needs to be diagnosed by solving the boundary value problem as described next.

\subsection{The diagnostic equations for boundary value problem}

To calculate $V$, the boundary value problem defined in section \ref{sec2.1} (by Eq. (\ref{eq2.1}) and Eq. (\ref{eq2.4})) will be solved when $\eta$ and $\phi$ are given; and boundary integral equations based on the Green’s theorem will be used. Next, the solving procedure will be briefed, which relies on two boundary integral equations: one is used to evaluate the velocity potential on the free surface $S$ and the other on the cylinder surface $C$.

\subsubsection{Evaluation point on the free surface}

For a field point on the free surface, the boundary integral equation is given as:

\begin{align}\label{eq2.7}
\begin{split}
\iint_{S \cup C}\left(\frac{1}{r}\right) \frac{\partial \Phi^{\prime}}{\partial n^{\prime}} \text{d} s^{\prime}=2 \pi \phi+\iint_{S \cup C} \Phi^{\prime} \frac{\partial}{\partial n^{\prime}}\left(\frac{1}{r}\right) \text{d} s^{\prime}.
\end{split}
\end{align}

The variables with a prime indicate those at source points $\left(\boldsymbol{x}^{\prime}, y^{\prime}\right)$, and those without a prime represent field points $\left(\boldsymbol{x}, y\right)$. The distance between a source point and a field point is $r=\sqrt{R^2+\left(y^{\prime}-y\right)^2}$ and the horizontal distance is $R=\left|\boldsymbol{x}^{\prime}-\boldsymbol{x}\right|$. After introducing $\text{d} s=\sqrt{1+\left|\nabla \eta\right|^2} \text{d} \boldsymbol{x}$, Eq. (\ref{eq2.7}) can be rewritten as:

\begin{align}\label{eq2.8}
\begin{split}
\iint_{S_0} \frac{V^{\prime}}{r} \text{d} \boldsymbol{x}^{\prime}=2 \pi \phi+\iint_{S_0} \phi^{\prime} \frac{\partial}{\partial n^{\prime}}\left(\frac{1}{r}\right) \sqrt{1+\left|\nabla^{\prime} \eta^{\prime}\right|^2} \text{d} \boldsymbol{x}^{\prime}+\iint_C \varphi^{\prime} \frac{\partial}{\partial n^{\prime}}\left(\frac{1}{r}\right) \text{d} s^{\prime}.
\end{split}
\end{align}

\noindent where $S_0$ is the projected area of $S$ to the horizontal plane. Similar with \cite{Wang2025}, the boundary integrals in Eq. (\ref{eq2.8}) involve two parts, i.e., one over the free surface and the other over the cylinder surface. They should be treated separately. 


For the integrals over the free surface, we refer to the work \cite{Wang2025,ClamondGrue2001,FructusEtAl2005} and introduce the following Taylor expansions about $(\eta_-/R)^2=0$:

\begin{align}\label{eq2.9}
\begin{split}
\frac{1}{r}=\sum_{j=0}^{(\mathcal{M}-1)/2} \frac{(-1)^j}{(2 j)!}\left[\eta^{2 j}_-\left(\nabla^{\prime} \cdot \nabla^{\prime}\right)^j \frac{1}{R}\right]+O\left(\epsilon^{\mathcal{M}}\right),
\end{split}
\end{align}

\begin{align}\label{eq2.10}
\begin{split}
\frac{\partial}{\partial n^{\prime}}\left(\frac{1}{r}\right) \sqrt{1+\left|\nabla^{\prime} \eta^{\prime}\right|^2}=\sum_{j=1}^{(\mathcal{M}-1)/2} \frac{(-1)^j}{(2 j-1)!}\nabla^{\prime} \cdot\left[\eta^{2j-1}_-\left(\nabla^{\prime} \cdot \nabla^{\prime}\right)^{j-1} \nabla^{\prime} \frac{1}{R}\right]+O\left(\epsilon^{\mathcal{M}}\right).
\end{split}
\end{align}

For the boundary integral over the cylinder surface, the following Taylor expansion about $\eta=0$ is adopted:

\begin{align}\label{eq2.11}
\begin{split}
\frac{\partial}{\partial n^{\prime}}\left(\frac{1}{r}\right)=- \frac{\boldsymbol{a}^\prime}{a} \cdot \nabla^{\prime}\left(\sum_{j=0}^{\mathcal{M}-1} \frac{(-1)^j}{j!} \eta^j \frac{\partial^j}{\partial y^{\prime j}} \frac{1}{r_{0}^{\prime}}\right)+O\left(\epsilon^{\mathcal{M}}\right),
\end{split}
\end{align}

\noindent where $\eta_-=\eta^{\prime}-\eta$, $\boldsymbol{a}^\prime/a=\left(\cos \theta^{\prime}, \sin \theta^{\prime}\right)$, $r_{0}^{\prime}=\sqrt{R^2+y^{\prime2}}$, $\epsilon=k_p|\eta|$ represent the magnitudes of wave steepness and $|\eta|$ is the characteristic wave amplitude (such as significant wave height). 

Next, we introduce the Fourier Transform on the $\boldsymbol{x}$ - (horizontal) plane:  
\begin{align}\label{eq2.12}
\begin{split}
    \mathcal{F} \left\{ * \right\} = \iint * e^{-\text{i} \boldsymbol{k}\cdot \boldsymbol{x}} \text{d} \boldsymbol{x} ,
\end{split}
\end{align}

\noindent where $\text{i}=\sqrt{-1}$, $\boldsymbol{k}=(k_1,k_2)$ denotes the wave number vector, $k=|\boldsymbol{k}|$, and $\mathcal{F}^{-1}\left\{ * \right\}$ denotes the inverse transform. Substituting the Taylor expansions of Eqs. (\ref{eq2.9} $\sim$ \ref{eq2.11}) into Eq. (\ref{eq2.8}) and applying Fourier Transform to both sides, while noting that:

\begin{align}\label{eq2.13}
\begin{split}
   \frac{1 }{ R} =2 \pi {\mathcal{F}}^{-1}\left\{ \frac{1}{k} e^{- \text{i} \boldsymbol{k} \cdot \boldsymbol{x}^{\prime}} \right\},
\end{split}
\end{align}
\begin{align}\label{eq2.14}
\begin{split}
    \frac{1 } {r_{0}^{\prime}} = 2 \pi {\mathcal{F}}^{-1}\left\{\frac{1}{ k}e^{-\text{i} \boldsymbol{k} \cdot \boldsymbol{x}^{\prime} - k |y^{\prime}|}  \right\},
\end{split}
\end{align}

\noindent a closed-form formulation for calculating $V$ can be derived:

\begin{align}\label{eq2.15}
\begin{split}
V=\sum_{j}^{\mathcal{M}} V_j+I_S.
\end{split}
\end{align}

Note that $ V_j$ stands for the quantity in terms of $j^{th}$ convolution and $I_S$ stands for the remaining integration term. The formula for $j=1$  is given as:

\begin{align}\label{eq2.16}
\begin{split}
\mathcal{F}\left\{V_1\right\}=k \mathcal{F}\{\phi\}+k \Psi,
\end{split}
\end{align}

The formulae for $j>1$ are given as:

\begin{align}\label{eq2.17}
\begin{split}
\mathcal{F}\{V_j\} &= -k \sum_{u=1}^j \frac{ (-1)^{u-1}(1-\langle j/2\rangle)}{(u-1)! (j-u)!} 
\Bigg[
\mathcal{F}\Bigl\{\eta^{u-1} \mathcal{F}^{-1}\bigl\{k^{j-3} \mathrm{i} \boldsymbol{k} \cdot 
\mathcal{F}\{\eta^{j-u} \nabla \phi\}\bigr\}\Bigr\} \\
& \quad + \mathcal{F}\Bigl\{\eta^{u-1} \mathcal{F}^{-1}\{k^{j-2} \chi_{j-u}\}\Bigr\}
\Bigg] + k \frac{(-1)^{j-1}}{(j-1)!} 
\mathcal{F}\Bigl\{\eta^{j-1} \mathcal{F}^{-1}\{k^{j-1} \Psi\}\Bigr\} \\
& - k \sum_{u=1}^j \frac{(-1)^{u-1}\langle j/2 \rangle}{(u-1)! (j-u)!}  
\mathcal{F}\Bigl\{\eta^{u-1} \mathcal{F}^{-1}\bigl\{k^{j-2} \mathcal{F}\{\eta^{j-u} V\}\bigr\}\Bigr\},
\end{split}
\end{align}

\noindent where

\begin{align}\label{eq2.18}
\begin{split}
\Psi(\boldsymbol{k})=-\frac{\partial}{\partial K} \int_0^{2 \pi} \int_{-\infty}^0 \psi^{\prime} e^{-\text{i} K \cos \left(\alpha-\theta^{\prime}\right)+k z^{\prime}} \text{d} z^{\prime} a  \text{d} \theta^{\prime},
\end{split}
\end{align}

\noindent and
\begin{align}\label{eq2.19}
\begin{split}
\chi_n(\boldsymbol{k})=-\frac{\partial}{\partial K} \int_L \varphi^{\prime} \eta^{\prime n}_C e^{-\text{i} K \cos \left(\alpha-\theta^{\prime}\right)} \text{d} l^{\prime}.
\end{split}
\end{align}

\noindent $\langle *\rangle$ is the remainder of integer $*$. $I_S$ is shown in Appendix \ref{appA}. Note that a surface-following vertical coordinate $z=y-\eta_C$ has been introduced, and $\varphi(y, \theta)=\psi(z, \theta)$, while $\eta_C=\eta(\rho=a)$ denotes the intersection of free surface and cylinder surface; $\alpha=\arctan(k_2/k_1)$, and $K=ka$.

\subsubsection{Evaluation point on the cylinder surface}

Following a similar procedure, for the field point on the cylinder surface, the boundary integral equation for evaluating $\varphi(y,\theta)$ can be given as:

\begin{align}\label{eq2.20}
\begin{split}
\iint_{S+C}\left(\frac{1}{r}\right) \frac{\partial \Phi^{\prime}}{\partial n^{\prime}} \text{d} s^{\prime}=2 \pi \varphi+\iint_{S+C} \Phi^{\prime} \frac{\partial}{\partial n^{\prime}}\left(\frac{1}{r}\right) \text{d} s^{\prime}.
\end{split}
\end{align}

After those integrals on the free surface being projected to horizontal plane, this equation can be reformulated as:

\begin{align}\label{eq2.21}
\begin{split}
2 \pi \varphi-\iint_{S_0} \frac{V^{\prime}}{r} \text{d} \boldsymbol{x}^{\prime}+\iint_{S_0} \phi^{\prime} \frac{\partial}{\partial n^{\prime}}\left(\frac{1}{r}\right) \sqrt{1+\left|\nabla^{\prime} \eta^{\prime}\right|^2} \text{d} \boldsymbol{x}^{\prime}+\iint_{\mathrm{C}} \varphi^{\prime} \frac{\partial}{\partial n^{\prime}}\left(\frac{1}{r}\right) \text{d} s^{\prime}=0.
\end{split}
\end{align}

For the integrals over the free surface, we introduce the following Taylor expansions about $\eta^{\prime}=0$:

\begin{align}\label{eq2.22}
\begin{split}
\frac{1}{r}=\sum_{j=0}^{\mathcal{M}-1} \frac{(-1)^j}{j!} \eta^{\prime j} \frac{\partial^j}{\partial y^j} \frac{1}{r_{0}}+O\left(\epsilon^{\mathcal{M}}\right),
\end{split}
\end{align}

\begin{align}\label{eq2.23}
\begin{split}
 \frac{\partial}{\partial n^{\prime}}\left(\frac{1}{r}\right) \sqrt{1+\left|\nabla^{\prime} \eta^{\prime}\right|^2} = 
 & -\frac{\partial}{\partial y} \frac{1}{r_{0}} +\sum_{j=1}^{(\mathcal{M}-1)/2} \frac{(-1)^j}{(2 j-1)!}\nabla^{\prime} \cdot\left[\eta^{\prime(2 j-1)}\left(\nabla^{\prime} \cdot \nabla^{\prime}\right)^{j-1} \nabla^{\prime} \frac{1}{r_{0}}\right] \\ 
& +\sum_{j=1}^{(\mathcal{M}-1)/2} \frac{(-1)^{j-1}}{(2 j)!}\nabla^{\prime} \cdot\left[\eta^{\prime(2 j)}\left(\nabla^{\prime} \cdot \nabla^{\prime}\right)^{j-1} \nabla^{\prime} \frac{\partial}{\partial y} \frac{1}{r_{0}}\right]+O\left(\epsilon^{\mathcal{M}}\right),
\end{split}
\end{align}

\noindent where $\boldsymbol{a}/a=\left(\cos \theta, \sin \theta\right)$, $r_{0}=\sqrt{|\boldsymbol{x}^{\prime}-\boldsymbol{a}|^2+y^2}$. For the boundary integral over the cylinder surface, there is:

\begin{align}\label{eq2.24}
\begin{split}
\frac{\partial}{\partial n^{\prime}}\left(\frac{1}{r}\right)=-\frac{\boldsymbol{a}^\prime}{a}\cdot \nabla^{\prime}\frac{1}{r_{a}},
\end{split}
\end{align}

\noindent where $r_{a}=\sqrt{|\boldsymbol{x}^{\prime}-\boldsymbol{a}|^2+\left(y^{\prime}-y\right)^2}$.  

Substitute Eqs. (\ref{eq2.22} $\sim$ \ref{eq2.24}) into Eq. (\ref{eq2.21}) and note that:

\begin{align}\label{eq2.25}
\begin{split}
    \frac{1}{r_{0}}=2 \pi {\mathcal{F}}_C^{-1}\left\{\frac{1}{ k}e^{-\text{i} \boldsymbol{k} \cdot \boldsymbol{x}^{\prime}-k| y|} \right\} ,
\end{split}
\end{align}

\begin{align}\label{eq2.26}
\begin{split}
    \frac{1} {r_{a}} =2 \pi {\mathcal{F}}_C^{-1}\left\{\frac{1 }{k}e^{-\text{i} \boldsymbol{k} \cdot \boldsymbol{x}^{\prime}-k\left|y^{\prime}-y\right|} \right\},
\end{split}
\end{align}

\noindent where $\mathcal{F}_C^{-1}$ is defined as:

\begin{align}\label{eq2.27}
\begin{split}
{\mathcal{F}}_C^{-1}\{*\}&= \frac{1}{(2 \pi)^2}\int \int * e^{\text{i} \boldsymbol{k}\cdot \boldsymbol{a}}\text{d} \boldsymbol{k} =\frac{1}{(2 \pi)^2} \int_0^{\infty} k \text{d} k \int_0^{2 \pi}  *  e^{\text{i} \boldsymbol{k} \cdot \boldsymbol{a}}\text{d} \alpha\\
&=\frac{1}{(2 \pi)^2} \int_0^{\infty} k \text{d} k  \int_0^{2 \pi}  *  e^{\text{i} K \cos (\alpha-\theta)}\text{d} \alpha\\
&=\frac{1}{(2 \pi)^2} \int \int *  e^{\text{i} K \cos (\alpha-\theta)}\text{d}k_1\text{d}k_2,
\end{split}
\end{align}

\noindent a closed-form formulation for calculating $\psi$ can be derived:

\begin{align}\label{eq2.28}
\begin{split}
\psi=\sum_{j}^{\mathcal{M}} \psi_j+I_C.
\end{split}
\end{align}

\noindent where $\psi_j$ stands for the quantity in terms of $j^{th}$ convolution and $I_C$ stands for the remaining integration term. The formula for $j=1$ is given as:

\begin{align}\label{eq2.29}
\begin{split}
& \psi_1-{\mathcal{F}}_C^{-1}\left\{\frac{\partial}{\partial K} \int_0^{2 \pi} \int_{-\infty}^0 \psi_1^{\prime} e^{-\text{i} K \cos \left(\alpha-\theta^{\prime}\right)} e^{-k\left|z^{\prime}-z\right|} \text{d} z^{\prime} a \text{d} \theta^{\prime}\right\} 
 \\& = {\mathcal{F}}_C^{-1}\left\{e^{k z}\left[k^{-1} {\mathcal{F}}\{V\}+{\mathcal{F}}\{\phi\}\right]\right\},
\end{split}
\end{align}

\noindent The formulas for $j>1$ are given as:

\begin{align}\label{eq2.30}
\begin{split}
& \psi_j - \mathcal{F}_C^{-1}\left\{ \frac{\partial}{\partial K} 
\int_0^{2\pi} \int_{-\infty}^{0} \psi_j^{\prime} 
e^{-\mathrm{i} K \cos(\alpha - \theta^{\prime})} 
e^{-k |z^{\prime} - z|} \, \mathrm{d}z^{\prime} \, a \, \mathrm{d}\theta^{\prime} \right\} \\
& = \frac{1}{(j-1)!} \mathcal{F}_C^{-1}\Bigg\{ 
e^{k z} \Bigg[ 
(-1)^{j-1} k^{j-2} \mathcal{F}\{\eta^{j-1} V\} + (-1)^j k^{j-3} \mathrm{i} \boldsymbol{k} \cdot \mathcal{F}\{\eta^{j-1} \nabla \phi\} + (-1)^j k^{j-2} \chi_{j-1} 
\Bigg] \Bigg\},
\end{split}
\end{align}

\noindent $I_C$ is shown in Appendix \ref{appA}.

\subsection{Successive approximations}\label{successive}

The Eq. (\ref{eq2.15}) and Eq. (\ref{eq2.29}) are implicit in the solution of $V$ and $\psi$, which means they depend on themselves during evaluation. In this section, a successive method is introduced to accelerate the computation. The convolution parts for $V$ and $\psi$ are rewritten as: 

\begin{align}\label{eq2.31}
\begin{split}
V=\sum_{l}^{\mathcal{M}} V_{(l)},
\end{split}
\end{align}

\begin{align}\label{eq2.32}
\begin{split}
\psi=\sum_{l}^{\mathcal{M}} \psi_{(l)}.
\end{split}
\end{align}

\noindent where $V_{(l)} \sim O\left(\epsilon^l\right)$ and $\psi_{(l)} \sim O\left(\epsilon^l\right)$. Substituting Eq. (\ref{eq2.31}) and Eq. (\ref{eq2.32}) into Eq. (\ref{eq2.15}) and Eq. (\ref{eq2.29}), and collecting terms of the same order in $\epsilon$ and their products, the formulations for $V_{(l)}$ and $\psi_{(l)}$ at each order can be derived successfully. The first order solution is given as:

\begin{align}\label{eq2.33}
\begin{split}
& \psi_{(1)}-{\mathcal{F}}_C^{-1}\left\{\frac{\partial}{\partial K} \int_0^{2 \pi} \int_{-\infty}^0 \psi_{(1)}^{\prime} e^{-\text{i} K \cos \left(\alpha-\theta^{\prime}\right)} e^{-k\left|z^{\prime}-z\right|} \text{d} z^{\prime} a \text{d} \theta^{\prime}\right\}  
 ={\mathcal{F}}_C^{-1}\left\{e^{k z}\left[k^{-1} {\mathcal{F}}\left\{V_{(1)}\right\}+{\mathcal{F}}\{\phi\}\right]\right\},
\end{split}
\end{align}

\begin{align}\label{eq2.34}
\begin{split}
{\mathcal{F}}\left\{V_{(1)}\right\}={k\mathcal{F}}\{\phi\}+k\Psi_{(1)},
\end{split}
\end{align}

\noindent where $\Psi_{(l)}$ denotes $\Psi$ composed of $\psi_{(l)}$. The following orders for $l=2\sim{\mathcal{M}}$ are given as:

\begin{align}\label{eq2.35}
\begin{split}
& \psi_{(l)} - \mathcal{F}_C^{-1}\left\{ \frac{\partial}{\partial K} 
\int_0^{2\pi} \int_{-\infty}^{0} \psi_{(l)}^{\prime} 
e^{-\mathrm{i} K \cos(\alpha - \theta^{\prime})} 
e^{-k |z^{\prime} - z|} \, \mathrm{d}z^{\prime} \, a \, \mathrm{d}\theta^{\prime} \right\} \\
& = \mathcal{F}_C^{-1}\Bigg\{ e^{k z} \Bigg[
\sum_{u=1}^{l} \frac{(-1)^{u-1}}{(u-1)!} k^{u-2} \mathcal{F}\{\eta^{u-1} V_{(l-u+1)}\} \\
& \quad + \frac{(-1)^{l}}{(l-1)!} k^{l-3} \mathrm{i} \boldsymbol{k} \cdot \mathcal{F}\{\eta^{l-1} \nabla \phi\}   + \sum_{u=1}^{l-1} \frac{(-1)^{u-1}}{u!} k^{u-1} \chi_{u(l-u)} 
\Bigg] \Bigg\},
\end{split}
\end{align}

\begin{align}\label{eq2.36}
\begin{split}
\mathcal{F}\left\{V_{(l)}\right\}=\sum_{u=1}^{l-1}-\frac{k^u}{u!} \mathcal{F}\left\{\eta^u V_{(l-u)}\right\}-\frac{k^{l-2}}{(l-1)!} \mathrm{i} \boldsymbol{k} \cdot \mathcal{F}\left\{\eta^{l-1} \nabla \phi\right\}+k \Im_{(l)}+k \Psi_{(l)}.
\end{split}
\end{align}

\noindent where

\begin{align}\label{eq2.37}
\begin{split}
\chi_{n(l)}(\boldsymbol{k})=-\frac{\partial}{\partial K} \int_L \psi_{(l)}^{\prime} \eta^{\prime n}_C e^{-\text{i} K \cos \left(\alpha-\theta^{\prime}\right)} \text{d} l^{\prime}.
\end{split}
\end{align}
\noindent $\Im_{(l)}$ for each order $l$ is given in Appendix \ref{appC}.

The equations for $V_{(l)}$ and $\psi_{(l)}$, e.g., Eq. (\ref{eq2.33}) and Eq. (\ref{eq2.34}), are two interdependent equations. By expressing $V_{(l)}$ in terms of $\psi_{(l)}$, we can isolate $\psi_{(l)}$ and solve it first. The formula for each order $l$ is given as:

\begin{align}\label{eq2.38}
\begin{split}
& \psi_{(l)} + \mathcal{F}_C^{-1}\left\{ \frac{\partial}{\partial K} 
\int_0^{2\pi} \int_{-\infty}^{0} \psi_{(l)}^{\prime} 
e^{-\mathrm{i} K \cos(\alpha - \theta^{\prime})} 
\left[ e^{k z + k z^{\prime}} - e^{-k |z^{\prime} - z|} \right] 
\, \mathrm{d}z^{\prime} \, a \, \mathrm{d}\theta^{\prime} \right\} = \mathcal{F}_C^{-1}\left\{ e^{k z} \xi_{(l)} \right\},
\end{split}
\end{align}

\noindent where $\xi_{(l)}$ is the convolution components for each order $l$. When $l=1$, $\xi_{(1)}=2\mathcal{F}\{\phi\}$. When $l=2 \sim{\mathcal{M}}$, $\xi_{(l)}$ is given as:

\begin{align}\label{eq2.39}
\begin{split}
& \xi_{(l)} = \sum_{u=1}^{l-1} \Bigg[ 
-2 \frac{\langle u/2 \rangle}{u!} k^{u-1} \mathcal{F}\{\eta^u V_{(l-u)}\} 
+ \frac{(-1)^{u-1}}{u!} k^{u-1} \chi_{u(l-u)} 
\Bigg]  -2 \frac{\langle l/2 \rangle}{(l-1)!} k^{l-3} \mathrm{i} \boldsymbol{k} \cdot 
\mathcal{F}\{\eta^{l-1} \nabla \phi\} 
+ \Im_{(l)}.
\end{split}
\end{align}

The solution methods for $\psi$ will be described in the next section. To be noted, the solution to this boundary value problem requires $\eta(T_n)$ and $\phi(T_n)$ being described at present time step $T = T_n$. Based on Eqs. (\ref{eq2.31}) \& (\ref{eq2.32}), $\psi_{(l)}$ and $V_{(l)}$ of different orders can be solved successively. Then, the diagnosed vertical surface velocity $V$ is substituted into the prognostic Eqs. (\ref{eq2.5}) \& (\ref{eq2.6}) to update $\eta$ and $\phi$ for the next time step $T = T_{n+1}$. The complete prognostic procedure will be elaborated in Part II.


\section{Solution to the velocity potential on cylinder surface}\label{sec3}

This section presents two approaches to solve Eq. (\ref{eq2.38}). 

\subsection{Fourier modes-based approach}

To calculate $V$, $\psi$ must first be solved as defined in Section \ref{successive}. It is assumed that the velocity potential on the cylinder surface $\psi_{(l)}$ can be explicitly expressed in terms of Fourier transform as:

\begin{align}\label{eq3.1}
\begin{split}
\psi_{(l)}(z, \theta)=\frac{1}{4\pi^2} \int_{-\infty}^{\infty} \int_{-\infty}^{\infty} B_{(l)} e^{\text{i}q^{\prime} z} e^{\text{i} m^{\prime} \theta} \text{d} q^{\prime} \text{d} m^{\prime} = \mathcal{F}_{\theta}^{-1}\left\{\mathcal{F}_{z}^{-1}\left\{B_{(l)}\right\} \right\},
\end{split}
\end{align}

\noindent where $B_{(l)}=B_{(l)}(q',m')$,  $q^{\prime}$ and $m^{\prime}$  are the wave numbers corresponding to $z$ and $\theta$, respectively, $\mathcal{F}_{z}^{-1}$ and $\mathcal{F}_{\theta}^{-1}$ are the corresponding inverse Fourier transforms. Note that the Fourier transforms for $z$ and $\theta$ maintain the same form as those defined for the horizontal plane (cf. Eq. \eqref{eq2.12}).

Now applying the Fourier transform $\mathcal{F}_{\theta}\{*\}(m)$ to both sides of Eq. (\ref{eq2.38}) will lead to:

\begin{align}\label{eq3.2}
\begin{split}
& \mathcal{F}_\theta\{\psi_{(l)}\} 
+ \mathcal{F}_\theta \Bigl\{ \mathcal{F}_C^{-1}\Bigl\{ \frac{\partial}{\partial K} 
\int_0^{2\pi} \int_{-\infty}^{0} \psi_{(l)}^{\prime} 
e^{-\mathrm{i} K \cos(\alpha-\theta^{\prime})} 
\bigl[e^{kz + kz^{\prime}} - e^{-k|z^{\prime}-z|}\bigr] 
\,dz^{\prime} a\, d\theta^{\prime} \Bigr\} \Bigr\} \\ 
   & = \mathcal{F}_\theta\bigl\{ \mathcal{F}_C^{-1}\{ e^{kz} \xi_{(l)} \} \bigr\}.
\end{split}
\end{align}

Next, we introduce the following Jacobi-Anger expansion using Bessel function of the first kind ($M$ = 1,2,3...)

\begin{align}\label{eq3.3}
\begin{split}
   & \mathcal{F}_{\theta} \left\{  e^{\text{i} K \cos (\alpha-\theta)} \right\} =\mathcal{F}_{\theta} \left\{  J_0(K)+2 \sum_{M=1}^{\infty} \text{i}^M J_M(K) \cos (M(\alpha-\theta)) \right\} \\ 
   & = e^{-\text{i}m\alpha}\int_0^{2\pi}  e^{\text{i} K \cos (\alpha-\theta)} e^{\text{i}m(\alpha -\theta) }\text{d} \theta  =2\pi \text{i}^m J_m(K)e^{-\text{i}m\alpha},
\end{split}
\end{align}

\noindent and

\begin{align}\label{eq3.4}
\begin{split}
&  \mathcal{F}_{\theta}\left\{\mathcal{F}_C^{-1} \left\{ * \right\}\right\} =\frac{\text{i}^m}{2\pi} \int_0^{\infty} k J_m(K) \int_0^{2\pi} * e^{-\text{i}m\alpha}\text{d}\alpha \text{d}k  =\frac{\text{i}^m}{2\pi} \int_{-\infty}^{\infty} \int_{-\infty}^{\infty} * J_m(K) e^{-\mathrm{i} m \alpha}\text{d} k_1 \text{d} k_2,
\end{split}
\end{align}

\noindent while noting that a mirror of $\psi_{(l)}$ about $z=0$ is created, therefore, Eq. (\ref{eq2.38}) can be reformulated as:

\begin{align}\label{eq3.5}
\begin{split}
& \int_{-\infty}^{\infty} B_{(l)} \cos (q' z) \, \mathrm{d} q' 
+ \frac{1}{2} \int_0^{\infty} K \frac{\partial J_m^2(K)}{\partial K} 
\int_{-\infty}^{\infty} B_{(l)} \int_{-\infty}^{0} \cos (q^{\prime} z^{\prime}) \left[ e^{k z + k z^{\prime}} - e^{-k |z^{\prime} - z|} \right] 
\, \mathrm{d} z^{\prime} \, \mathrm{d} q^{\prime} \, \mathrm{d} k \\
& = \mathrm{i}^m \int_{-\infty}^{\infty} \int_{-\infty}^{\infty} 
e^{k z} \xi_{(l)} J_m(K) e^{-\mathrm{i} m \alpha} \, \mathrm{d} k_1 \, \mathrm{d} k_2.
\end{split}
\end{align}

Next, applying transform $\mathcal{F}_{z}\{*\}(q)$ to Eq. (\ref{eq3.2}) while noting that

\begin{align}\label{eq3.6}
\begin{split}
\begin{aligned} \mathcal{F}_{z}\left\{\int_{-\infty}^0 \cos \left(q^{\prime} z^{\prime}\right)\left[e^{k z+k z^{\prime}}-e^{-k\left|z^{\prime}-z\right|}\right] \text{d} z^{\prime}\right\}
=\frac{2 k}{k^2+q^{\prime 2}}\left[\frac{k}{k^2+q^2}-\delta(q^{\prime}-q)\right],
\end{aligned}
\end{split}
\end{align}

\noindent the following equation to be solved for $B_{(l)}$ can be given as:

\begin{align}\label{eq3.7}
\begin{split}
&\int_{-\infty}^{\infty} B_{(l)} \left\{ \delta \left [ 1-\int_0^{\infty}\frac{k^2a}{k^2+q^{\prime 2}} \frac{\partial J_m^2(K)}{\partial K}  \text{d} k \right] +\int_0^{\infty} \frac{k^3a}{(k^2+q^2)(k^2+q^{\prime 2})} \frac{\partial J_m^2(K)}{\partial K} \text{d} k \right\}\text{d}q^{\prime} \\& = {\mathrm{i}^m} \int_{-\infty}^{\infty} \int_{-\infty}^{\infty}  \frac{k}{k^2+q^2} \xi_{(l)}J_m(K) e^{-\mathrm{i} m \alpha}\text{d} k_1 \text{d} k_2,
\end{split}
\end{align}

\noindent where $\delta$ is the Dirac Delta function, the calculation will be introduced in the next section.

\subsection{Legendre polynomials-based approach}

This subsection provides an alternative approach to calculate $\psi_{(l)}$ using Legendre transform to depict its variation in vertical direction using a series of Legendre polynomials. This is feasible as the wave-induced motions are mainly concentrated near the free surface. Unlike the Fourier transform providing solutions on a uniform grid, we introduce the following vertical coordinate by transforming $z \in (-\infty,0 ]$ to $\zeta \in [-1,1]$ via:

\begin{align}\label{eq3.8}
\begin{split}
\zeta=2 e^z-1.
\end{split}
\end{align}

In this sense, the velocity potential on cylinder surface $\psi_{(l)}$ can be re-expressed as:

\begin{align}\label{eq3.9}
\begin{split}
\psi_{(l)}(\zeta, \theta)=\frac{1}{2\pi}\int_{-\infty}^{\infty} \left[\sum_{N^{\prime}=0}^{\infty} \frac{2 N^{\prime}+1}{2} P_{N^{\prime}}\left(\zeta\right) L_{(l)}\right] e^{\mathrm{i} m^{\prime} \theta}  \text{d} m^{\prime} = \mathcal{F}_{\theta}^{-1}\left\{\mathcal{J}_{\zeta}^{-1}\left\{L_{(l)}\right\} \right\},
\end{split}
\end{align}

\noindent where $P_{N^{\prime}}(\zeta)$ is Legendre polynomial at order $N^{\prime}$ = 0,1,2... :

\begin{align}\label{eq3.10}
P_0(\zeta)=1, P_1(\zeta)=\zeta, P_2(\zeta)=\frac{1}{2}\left(3 \zeta^2-1\right) \ldots 
\end{align}

\noindent and $L_{(l)}=L_{(l)}(N^{\prime}, m^{\prime})$ is the corresponding coefficient to be computed,  $N^{\prime}$ is the Legendre polynomial order corresponding to $\zeta$, $\mathcal{J}_{\zeta}^{-1}$ is the corresponding inverse Legendre transform. The Legendre transform for $\zeta$ - coordinate is given as:

\begin{align}\label{eq3.11}
\begin{split}
\mathcal{J}_\zeta\{*\}=\int_{-1}^1 * P_{N^{\prime}}(\zeta) \text{d} \zeta.
\end{split}
\end{align}

Now applying the Fourier transform $\mathcal{F}_{\theta}\{*\}(m)$ to both sides of Eq. (\ref{eq2.38}). In analogy to Eq. (\ref{eq3.2}), the following equation can be derived:

\begin{align}\label{eq3.12}
\begin{split}
& \sum_{N=0}^{\infty} \frac{2 N+1}{2} P_N(\zeta) L_{(l)}+\frac{1}{2} \sum_{N^{\prime}=0}^{\infty} \frac{2 N^{\prime}+1}{2} L_{(l)} \int_0^{\infty} K \frac{\partial J_m^2(K)}{\partial K}\int_{-1}^1 \mathcal{E} P_{N^{\prime}}\left(\zeta^{\prime}\right) \text{d} \zeta^{\prime} \text{d} k  
\\& =\frac{\mathrm{i}^m}{2 \pi} \int_{-\infty}^{\infty} \int_{-\infty}^{\infty}\left(\frac{\zeta+1}{2}\right)^k \xi_{(l)} J_m(K) e^{-\mathrm{i} m \alpha} \text{d} k_1 \text{d} k_2,
\end{split}
\end{align}

\noindent where 

\begin{align}\label{eq3.13}
\mathcal{E}(k,\zeta^\prime,\zeta)=\frac{e^{k \ln \left[\left(\zeta^{\prime}+1\right)\left(\zeta+1\right) / 4\right]}-e^{-k\left|\ln \left[\left(\zeta^{\prime}+1\right) /\left(\zeta+1\right)\right]\right|}} {\zeta^{\prime}+1}.
\end{align}

\noindent is a function arising from the coordinate transformation, which will be further discussed subsequently.

Next, applying Legendre transform $\mathcal{J}_{\zeta}\{*\}(N)$ to Eq. (\ref{eq3.12}) while noting that

\begin{align}\label{eq3.14}
\begin{split}
\int_{-1}^1\left(\frac{\zeta+1}{2}\right)^k P_N(\zeta) \text {d}\zeta=\left\{
\begin{array}{c}\frac{2}{k+1}, 
\text { when } N=0, 
\\ \frac{2 \prod_{n=0}^{N-1}(k-n)}{\prod_{n=0}^{N}(k+n+1)}, 
\text { when } N>0 ;
\end{array}\right.
\end{split}
\end{align}

\noindent and the analytical forms of this dual Legendre transforms $\int_{-1}^1 \int_{-1}^1 \mathcal{E}  P_N(\zeta)P_{N^{\prime}}\left(\zeta^{\prime}\right)  \text{d} \zeta\text{d} \zeta^{\prime}$ for $N^{\prime}$ and $N$ equal to $0,1,2$ are summarized in table \ref{tab1}, the following equation to be solved for $L_{(l)}$ can be given as:

\begin{align}\label{eq3.15}
\begin{split}
&\sum_{N^{\prime}=0}^{\infty}L_{(l)} \left\{\frac{2 N^{\prime}+1}{4}  \int_0^{\infty} K \frac{\partial J_m^2(K)}{\partial K} \int_{-1}^1 \int_{-1}^1 \mathcal{E} P_N(\zeta)P_{N^{\prime}}\left(\zeta^{\prime}\right) \text{d}\zeta \text{d} \zeta^{\prime}\text{d} k +I_{N^{\prime},N}\right\}  \\& =\frac{\mathrm{i}^m}{2 \pi} \int_{-\infty}^{\infty} \int_{-\infty}^{\infty} \int_{-1}^1\left(\frac{\zeta+1}{2}\right)^k P_N(\zeta)  \xi_{(l)} J_m(K) e^{-\mathrm{i} m \alpha} \text{d} \zeta\text{d} k_1 \text{d} k_2,
\end{split}
\end{align}

\noindent where $I_{N^{\prime},N}$ is equal to 1 when $N^{\prime}=N$, otherwise $I_{N^{\prime},N}$ is equal to 0.

\begin{table}[width=.9\linewidth,cols=4,pos=h]
\caption{Analytical Expressions for $\int_{-1}^1 \int_{-1}^1 \mathcal{E} P_N(\zeta)P_{N^{\prime}}\left(\zeta^{\prime}\right) \text{d} \zeta\text{d} \zeta^{\prime}$}\label{tab1}
\begin{tabular*}{\tblwidth}{@{} C C C C @{}}
\toprule
Legendre-orders & $N=0$ & $N=1$ & $N=2$ \\
\midrule
$N^\prime=0$ & $\displaystyle \frac{-4}{k+1}$ 
             & $\displaystyle \frac{4}{(k+2)(k+1)}$ 
             & $\displaystyle \frac{4(k-1)}{(k+3)(k+2)(k+1)}$ \\
             
$N^\prime=1$ & $\displaystyle \frac{4}{(k+1)^2}$ 
             & $\displaystyle \frac{2(k-1)}{(k+2)(k+1)^2}-\frac{2}{3}\left(\frac{1}{k+1}+\frac{1}{k+2}\right)$ 
             & $\displaystyle \frac{4\left(k^2+1\right)}{(k+3)(k+2)(k+1)^2}$ \\
             
$N^\prime=2$ & $\displaystyle \frac{4(k-2)}{(k+2)(k+1)^2}$ 
             & $\displaystyle \frac{4\left(k^2+2\right)}{(k+2)^2(k+1)^2}$ 
             & $\displaystyle \frac{2(k-1)^2(k-2)}{(k+3)(k+2)^2(k+1)^2}-\frac{2}{5} \left(\frac{1}{k+2}+\frac{1}{k+3}\right)$ \\
\bottomrule
\end{tabular*}
\end{table}

In summary, two approaches have been described in this section to compute $\psi_{(l)}$, i.e., Fourier mode- and Legendre polynomials-based approaches. In the next, they will be verified against the linear analytical solution for a regular wave scattering problem.

\section{Model Discretizations}\label{sec4}

In this section, we introduce the solution for variables $B_{(l)}(q,m)$ and $L_{(l)}(N,m)$ in discrete forms.

\subsection{Linear equations}\label{sec4.1}

For the Fourier modes-based approach, assume the distribution of $\psi_{(l)}(z,\theta)$ in the vertical coordinate $z$ is fitted by $N_q$ Fourier modes, the wavenumber $q=0,\Delta q, 2\Delta q,...,(N_q-1)\Delta q$. The analysis interval for vertical coordinate $z$ is $\left[-h,h\right]$, where $h$ is the maximum depth of interest, which is assumed to be deep enough so that the cylinder surface velocity potential reduces to 0. The $\psi_{(l)}(z,\theta)$ above $z=0$ is set to be 0. The wavenumber $q$ is mapped to the interval $\left[-\pi,\pi\right]$ by setting the discrete step as $\Delta q=\pi/h$. Assume the distribution of $\psi_{(l)}(z,\theta)$ in the angular coordinate $\theta$ is fitted by $N_m$ Fourier modes, the wavenumber $m=0,1,2,...,N_m-1$. Then, the solution for $B_{(l)}(q,m)$ is discretized into the solution of $N_q$ linear equations for a fixed wavenumber $m$, and Eq. (\ref{eq3.7}) is changed to:

\begin{align}\label{eq4.1}
\begin{split}
[A]_{N_q \times N_q}\left\{B_{(l)} \right\}_{N_q}=\{C_{(l)}\}_{N_q},
\end{split}
\end{align}

\noindent where

\begin{align}\label{eq4.2}
\begin{split}
\delta=\int_{-h}^0 \cos (q z) \cos \left(q^{\prime} z\right) d z=\left\{\begin{array}{cl}h, & \left(q=q^{\prime}\right)=0, \\ 0.5 h, & \left(q=q^{\prime}\right) \neq 0, \\ 0, & \left(q \neq q^{\prime}\right).\end{array}\right.
\end{split}
\end{align}

For the Legendre polynomials-based approach, assume the distribution of $\psi_{(l)}(\zeta,\theta)$ in the vertical coordinate $\zeta$ is fitted by $N_l$ Legendre polynomials, the polynomial order $N=0,1,2,…,N_l-1$. The angular distribution is also assumed to be fitted by $N_m$ Fourier modes, with the wavenumber $m=0,1,2,…,N_m-1$. Then, the solution for $L_{(l)}(N,m)$  is discretized into the solution of $N_l$ linear equations for a fixed wavenumber $m$, and Eq. (\ref{eq3.15}) is changed to:

\begin{align}\label{eq4.3}
[D]_{N_l \times N_l} \left\{L_{(l)} \right\}_{N_l}=\{E_{(l)}\}_{N_l}.
\end{align}

\noindent It should be noted that the calculated wavenumbers are all greater than zero. This is because $B_{(l)}(-q,m)$ and $B_{(l)}(q,m)$ are equal due to its even extension characteristics, $B_{(l)}(q,-m)$ $\&$ $B_{(l)}(q,m)$ and $L_{(l)}(N,-m)$ $\&$ $L_{(l)}(N,m)$ are conjugates.

\subsection{Matrix computation}

Both matrices $[A]$ and $[D]$ are integrals with respect to the variable $k$. Ensuring convergence of the integrals is crucial for ensuring correct matrix computation. Analyzing the components of the integrands in the matrices (table \ref{tab2}), it can be concluded that integrands with convergence of $O(1/k)$ converge quickly as $k$ increases and require no further discussion. However, those with convergence of $O(1)$ require further discussion of the accuracy of the integrals. In table \ref{tab2}, $C_{N,N^{\prime}}=-4/(2N+1)$; $k_t$ refers to a sufficiently large $k$ that the integrand component reaches the approximation value. By comparing the difference between the integrand component and the approximation value, we define the error $\text{Err}(k)$ as:

\begin{align}\label{eq4.4}
\begin{split}
&\text{Err}(k)=\frac{\left|\frac{k^2 a}{k^2+q^{\prime 2}}-a\right|}{|a|} \times 100 \% \quad \text{for}\quad [A], \\
&\text{Err}(k)=\frac{\left|k a \int_{-1}^1 \int_{-1}^1 \mathcal{E} P_N(\zeta) P_{N^{\prime}}\left(\zeta^{\prime}\right) \mathrm{d} \zeta \mathrm{d} \zeta^{\prime}-a C_{N, N^{\prime}}\right|}{\left|a C_{N, N^{\prime}}\right|} \times 100 \% \quad \text{for}\quad [D].
\end{split}
\end{align}

\begin{table}[width=.9\linewidth,cols=4,pos=h]
\caption{Integrand components in matrices}\label{tab2}
\begin{tabular*}{\tblwidth}{@{} C C C C @{}}
\toprule
Matrix & Integrand components & Approximation value at $k_t$ & Convergence \\
\midrule
$[A]$ & $\displaystyle \frac{k^2 a}{k^2+q^{\prime 2}}$ 
      & $a$ 
      & $O(1)$ \\

$[A]$ & $\displaystyle \frac{k^3 a}{\left(k^2+q^2\right)\left(k^2+q^{\prime 2}\right)}$ 
      & $\displaystyle \frac{a}{k}$ 
      & $\displaystyle O\left(\frac{1}{k}\right)$ \\

$[D]$ when $N=N^\prime$ 
      & $\displaystyle k a \int_{-1}^1 \int_{-1}^1 \mathcal{E} P_N(\zeta) P_{N^{\prime}}\left(\zeta^{\prime}\right) \mathrm{d} \zeta \mathrm{d} \zeta^{\prime}$ 
      & $a C_{N, N^{\prime}}$ 
      & $O(1)$ \\

$[D]$ when $N \neq N^\prime$ 
      & $\displaystyle k a \int_{-1}^1 \int_{-1}^1 \mathcal{E} P_N(\zeta) P_{N^{\prime}}\left(\zeta^{\prime}\right) \mathrm{d} \zeta \mathrm{d} \zeta^{\prime}$ 
      & $\displaystyle \frac{a}{k}$ 
      & $\displaystyle O\left(\frac{1}{k}\right)$ \\

$[A]$ \& $[D]$ 
      & $\displaystyle \frac{\partial J_m^2(ka)}{\partial K}$ 
      & $\displaystyle -\frac{2}{\pi K} \sin \left(2 K-m \pi-\frac{\pi}{2}\right)$ 
      & -- \\
\bottomrule
\end{tabular*}
\end{table}

\begin{figure}
	\centering
	\begin{subfigure}[b]{0.48\textwidth}
		\centering
		\includegraphics[width=\textwidth]{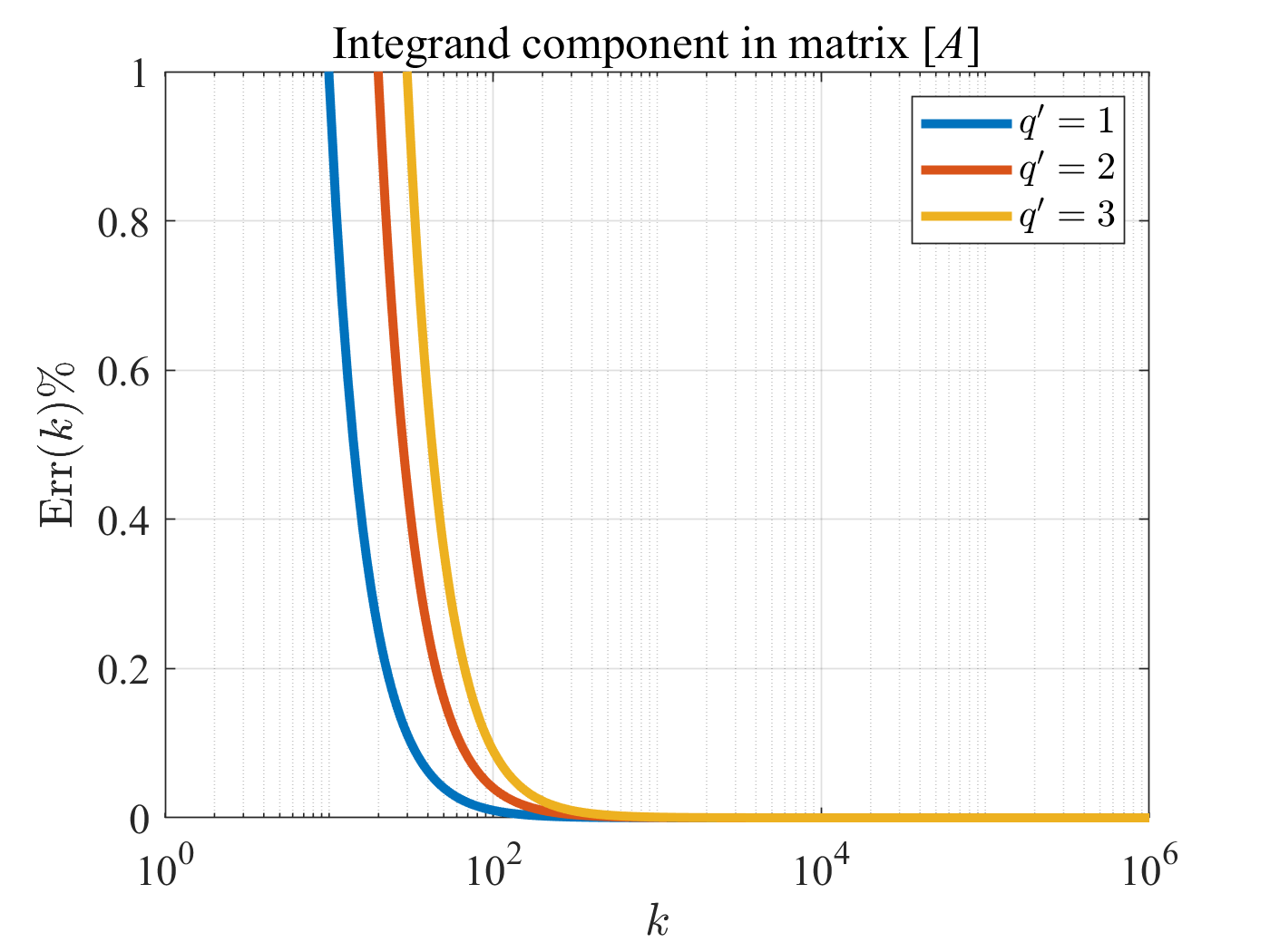}
	\end{subfigure}
	\hfill
	\begin{subfigure}[b]{0.48\textwidth}
		\centering
		\includegraphics[width=\textwidth]{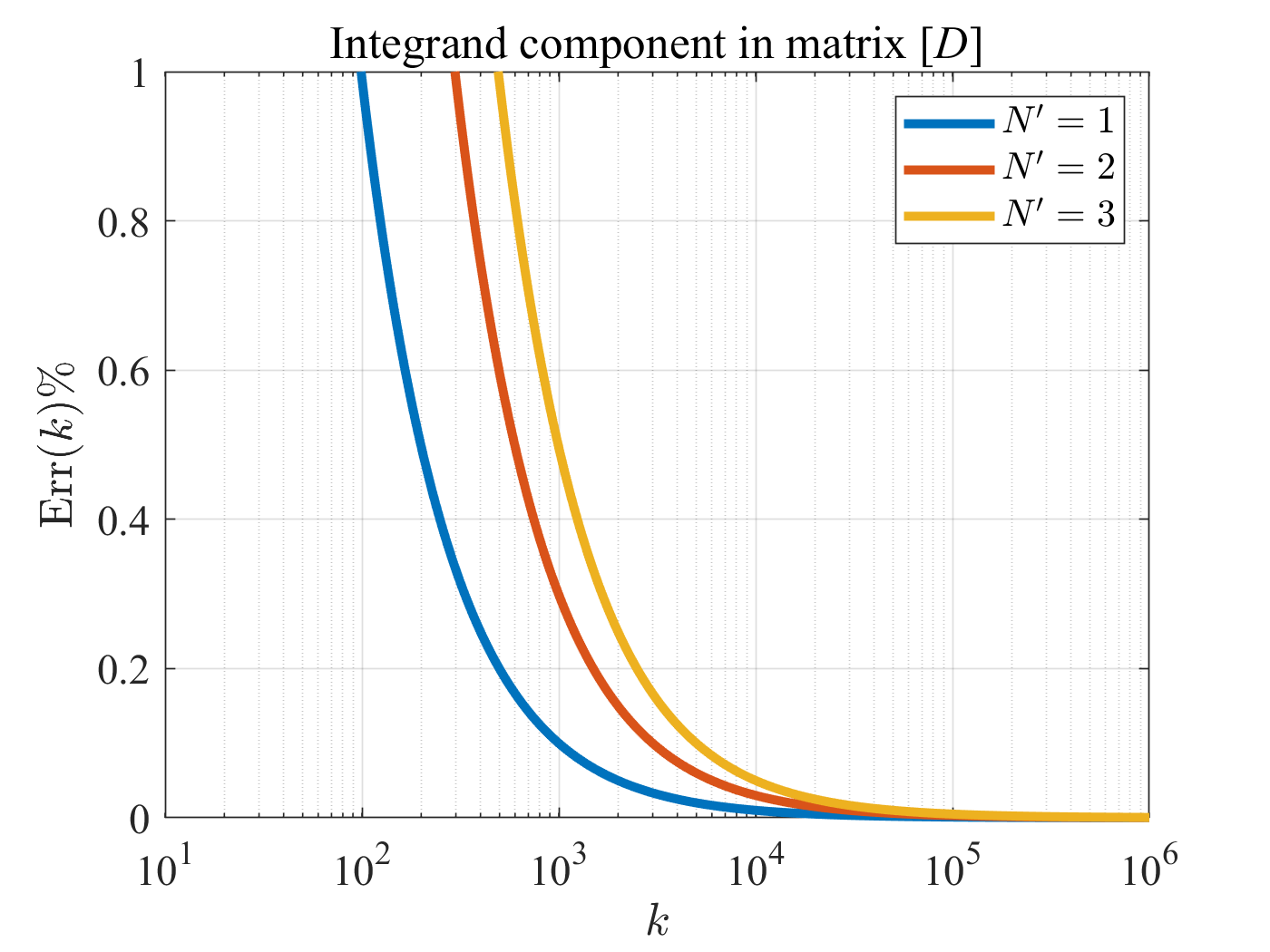}
	\end{subfigure}
	\caption{Change of $\text{Err}(k)$ with respect to $k$ for different $q^\prime$ and $N^\prime$.}
	\label{fig2}
\end{figure}

Figure \ref{fig2} shows the change of $\text{Err}(k)$ with respect to $k$ for different $q^{\prime}$ and $N^{\prime}$. It can be seen that for any $k \geq  5 \times 10^4$, the error is small enough (at the level of $10^{-2}\%$) to allow the integrand components to be close to $a$ and $a C_{N, N^{\prime}}$. Thus, $k_t$ is set to be $5 \times 10^4$ in practical computations. 

Since the integral $\int {\partial J_m^2(K)}/{\partial K} \text{d} k$ yields $J_m^2(K) / a+ \text{constant}$ for large $k$, the integral part of matrix $[A]$ and matrix $[D]$ with convergence of $O(1)$ can be rewritten as the superposition of a integral term $I$ from 0 to $k_t$ and a constant term, which saves the computational effort led by the oscillation frequency of the integrands at large $k$: 

\begin{align}\label{eq4.5}
\begin{split}
& \int_0^{\infty} \frac{\partial J_m^2(K)}{\partial K} \frac{k^2 a}{k^2+q^{\prime 2}} \text{d} k=\int_0^{k_t} \frac{\partial J_m^2(K)}{\partial K} \frac{k^2 a}{k^2+q^{\prime 2}} \text{d} k-J_m^2\left(k_t a\right) \;\text{for} \; [A],
\\ & \int_0^{\infty} \frac{\partial J_m^2(K)}{\partial K} k a \int_{-1}^1 \int_{-1}^1 \mathcal{E} P_N(\zeta) P_{N^{\prime}}\left(\zeta^{\prime}\right) \text{d} \zeta \text{d} \zeta^{\prime} \text{d} k 
\\ & =\int_0^{k_t} \frac{\partial J_m^2(K)}{\partial K} k a \int_{-1}^1 \int_{-1}^1 \mathcal{E} P_N(\zeta) P_{N^{\prime}}\left(\zeta^{\prime}\right) \text{d} \zeta \text{d} \zeta^{\prime} \text{d} k-C_{N, N^{\prime}} J_m^2\left(k_t a\right) \;\text{for}\;[D].
\end{split}
\end{align}

The influence of $\Delta k$ in the integral term is also discussed. A sufficiently small $\Delta k$ is required to make the integral converge. Assuming that the smallest $\Delta k_{\text{min}}= 10^{-3}$ gets the most accurate integral value, define the error $\text{Err}(\Delta k)$ as:

\begin{align}\label{eq4.6}
\begin{split}
\text{Err}(\Delta k)=\frac{\left|I(\Delta k)-I\left(\Delta  k_{\text{min}}\right)\right|}{\left|I\left(\Delta  k_{\text{min} }\right)\right|} \times 100 \%.
\end{split}
\end{align}

\begin{figure}
	\centering
	\begin{subfigure}[b]{0.48\textwidth}
		\centering
		\includegraphics[width=\textwidth]{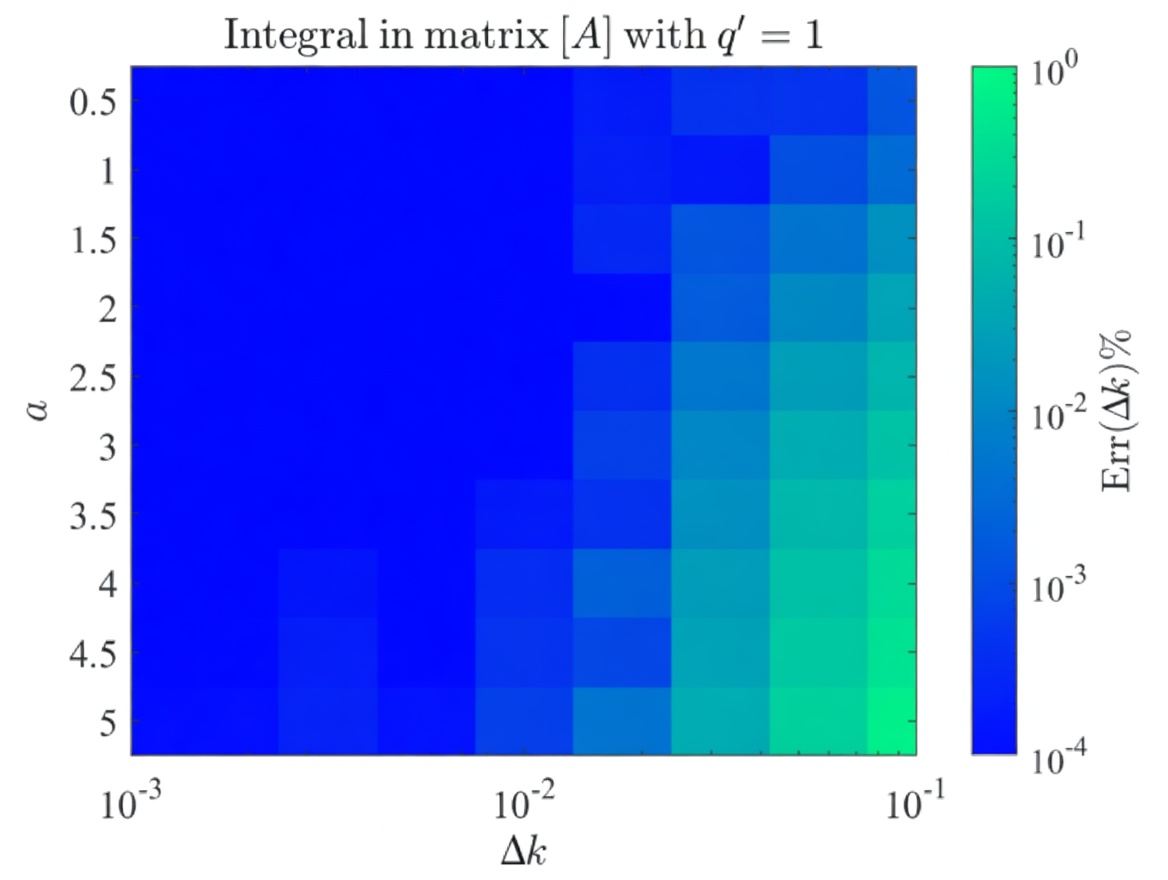}
	\end{subfigure}
	\hfill
	\begin{subfigure}[b]{0.48\textwidth}
		\centering
		\includegraphics[width=\textwidth]{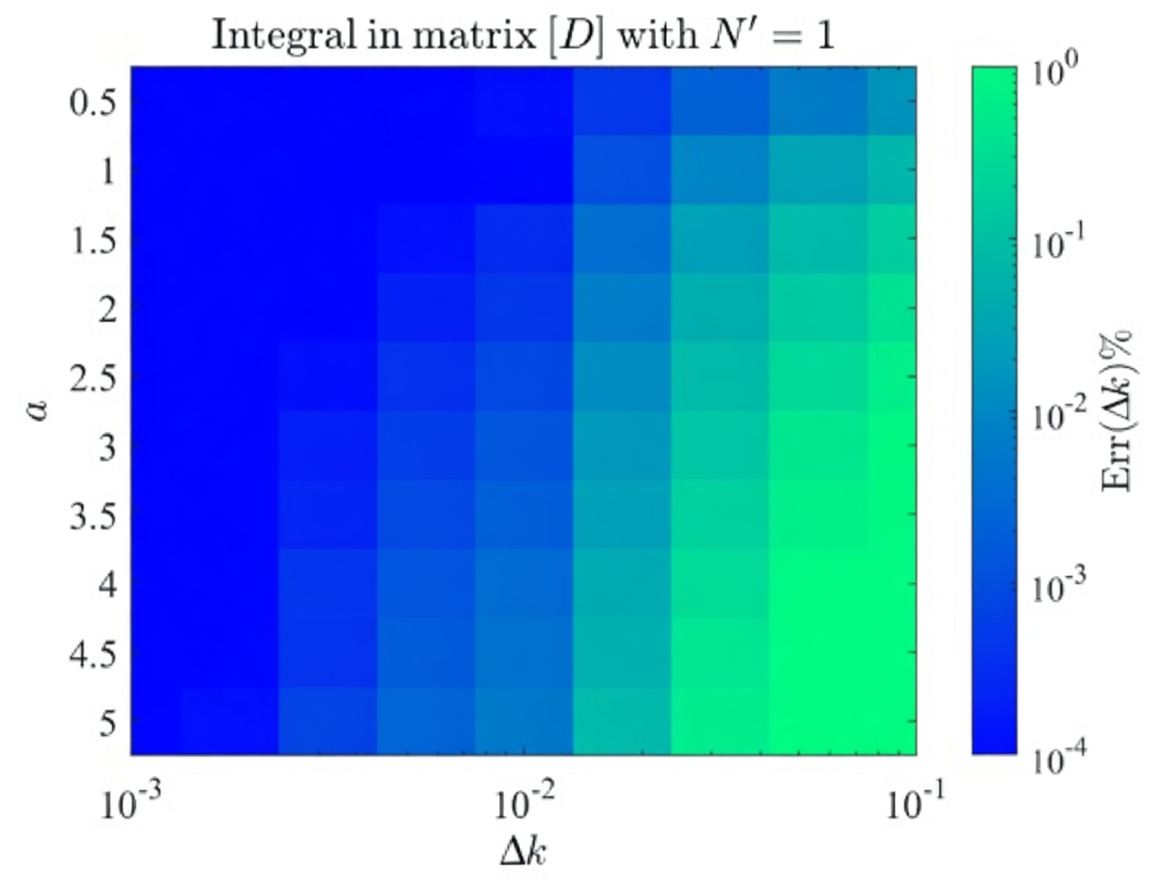}
	\end{subfigure}
	\caption{Change of $\text{Err}(\Delta k)$ with respect to $m=0$, different $\Delta k$ and different cylinder radius a when $q^\prime=N^\prime=1$: $[A]$ and $[D]$.}
	\label{fig3}
\end{figure}

The integral $I$ is calculated using the 16-point Gauss quadrature method. Figure \ref{fig3} shows the change of $\text{Err}(\Delta k)$ with respect to $m=0$, different $\Delta k$ and different cylinder radius $a$ when $q^{\prime}=N^{\prime}=1$.  It can be seen from figure \ref{fig3}, a smaller $\Delta k$ is required as the cylinder radius increases. In general, setting $\Delta k= 10^{-2}$ and $\Delta k= 10^{-1}$ can ensure $\text{Err}(\Delta k)$ is within $10^{-2}\%$ for different $m$ and $a$ when $k_t=10^4$ for the integral $I$ in both matrix $[A]$ and matrix $[D]$, respectively. 

Finally, different values of $k_t$ are prescribed to examine the convergence of the integral in Eq. (\ref{eq4.5}), with $\Delta k = 0.01$ and $m = 0$. As presented in Table \ref{tab3}, setting $k_t = 5 \times 10^4$ ensures that the overall integration error is controlled at the order of $10^{-4}$. That being said, it can now be ensured that the matrix calculations have reached convergence.

\begin{table}
\caption{Convergence of integral quantities ($\times 10^{-1}$) in Eq. (\ref{eq4.5}) with respect to the truncation parameter $k_t$}\label{tab3}
\renewcommand{\arraystretch}{1.2} 
\begin{tabular*}{\textwidth}{@{\extracolsep{\fill}} C C C C C C C @{}}
\toprule
$k_t$ & $I_{\mathrm{[A]}}, q^\prime = 1$ & $I_{\mathrm{[A]}}, q^\prime = 2$ & $I_{\mathrm{[A]}}, q^\prime = 3$ & $I_{\mathrm{[D]}}, N^\prime = 1$ & $I_{\mathrm{[D]}}, N^\prime = 2$ & $I_{\mathrm{[D]}}, N^\prime = 3$ \\
\midrule
$1.0 \times 10^3$ & $-5.24108$ & $-2.75345$ & $-1.75971$ & $5.19281$ & $4.34164$ & $3.15859$ \\
$2.5 \times 10^3$ & $-5.24108$ & $-2.75345$ & $-1.75971$ & $5.17443$ & $4.33958$ & $3.15979$ \\
$5.0 \times 10^3$ & $-5.24108$ & $-2.75345$ & $-1.75971$ & $5.17570$ & $4.33972$ & $3.15970$ \\
$7.5 \times 10^3$ & $-5.24108$ & $-2.75345$ & $-1.75971$ & $5.17679$ & $4.33984$ & $3.15963$ \\
$1.0 \times 10^4$ & $-5.24108$ & $-2.75345$ & $-1.75971$ & $5.17589$ & $4.33974$ & $3.15969$ \\
$2.5 \times 10^4$ & $-5.24108$ & $-2.75345$ & $-1.75971$ & $5.17438$ & $4.33958$ & $3.15979$ \\
$5.0 \times 10^4$ & $-5.24108$ & $-2.75345$ & $-1.75971$ & $5.17458$ & $4.33960$ & $3.15978$ \\
$7.5 \times 10^4$ & $-5.24108$ & $-2.75345$ & $-1.75971$ & $5.17463$ & $4.33960$ & $3.15978$ \\
$1.0 \times 10^5$ & $-5.24108$ & $-2.75345$ & $-1.75971$ & $5.17447$ & $4.33959$ & $3.15979$ \\
\bottomrule
\end{tabular*}
\end{table}


\section{Results and discussions}\label{sec5}
In this chapter, the newly developed diagnosis solver is systematically verified.The verification is performed in two stages. First, the convergence properties of the proposed formulations are examined with respect to the Fourier/Legendre spectral truncation parameters $(N_q,N_m)$ or $(N_l,N_m)$, together with the spatial resolution of the computational domain. Second, the fully converged numerical solutions are directly compared with the corresponding analytical solutions.

To provide a rigorous verification under well-controlled conditions, steady-state regular wave–cylinder interactions with weakly nonlinear effects are considered. A small wave steepness $A=0.01$ is selected, and different scattering parameters $a=0.5,1,3$, and $5$ are investigated, covering a broad range from the long-wave regime ($a<1$) to the short-wave and strongly diffracting regime ($a>1$). Here, $A$ and $a$ denote the dimensionless incident wave amplitude and cylinder radius, respectively, defined as the physical incident wave amplitude and cylinder radius multiplied by the incident wavenumber.

For the verification, the free surface boundary conditions are prescribed using the first-order diffraction solution of Havelock (1940)\cite{Havelock1940}. Based on these prescribed boundary conditions, the first- and second-order cylinder surface velocity potentials (given by first- and second-order convolutions) are calculated using the proposed Legendre-based formulations and compared with the analytical solutions given by Havelock (1940) and the Fourier-based formulations. It should be noted that only first-order incident-wave effects are included in the prescribed free-surface boundary conditions, while the second-order potential is taken into account on the cylinder surface for verification (reflecting weak nonlinear quadratic convolution effects).

\subsection{First-order convolutions}
The Fourier/Legendre series expansions for $B_{\left(l\right)}$ and $L_{\left(l\right)}$ are truncated at spectral truncation parameters $(N_q,N_m)$ or $(N_l,N_m)$ when computing the cylinder surface velocity potential $\psi_{\left(l\right)}$. Here, $N_q$ and $N_l$ governs the discretization of the spectrum in the vertical coordinate $z$, while $N_m$ controls the resolution in the azimuthal coordinate $\theta$, which have been defined in section \ref{sec4.1}. Based on findings from our previous work \cite{ShangChenLinWang2026}, the parameters $(N_q,N_m)=(2^{10},2^6)$ are adopted for all subsequent computations to ensure accurate results for $\psi_{\left(l\right)}$ using the Fourier-based approach.

A key limitation of the Fourier-based method is the considerable computational cost incurred by large matrices (or a large number of $N_q$), especially during matrix inversion, assembly, and multiplication operations in realistic simulations. This is also a primary challenge of the BEM. In BEM applications to wave-cylinder interactions, the discretization of the fluid boundary (including the free surface and cylinder surfaces) generates a fully dense matrix whose size scales with the square of the number of panels $N_p$. Solving the resulting linear system typically requires $O(N_p^3)$ time (or at least $O(N_p^2)$ with iterative solvers), leading to prohibitive computational time and memory demands when high-resolution meshes or multiple cylinders are involved in practical 3D simulations \citep{TengGou2017,NokobYeung2020}.

\begin{figure}
	\centering
	\begin{subfigure}[b]{0.48\textwidth}
		\centering
		\includegraphics[width=\textwidth]{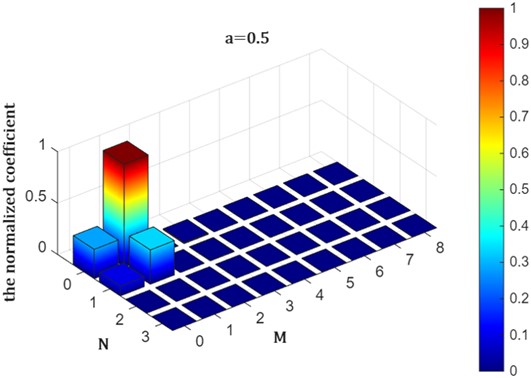}
	\end{subfigure}
	\hfill
	\begin{subfigure}[b]{0.48\textwidth}
		\centering
		\includegraphics[width=\textwidth]{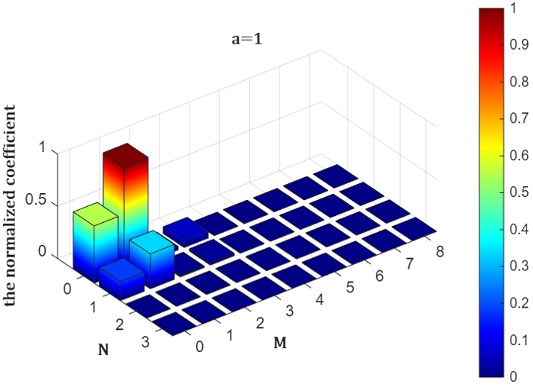}
	\end{subfigure}
	\begin{subfigure}[b]{0.48\textwidth}
		\centering
		\includegraphics[width=\textwidth]{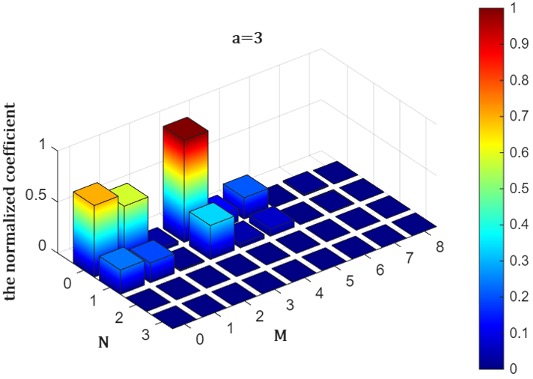}
	\end{subfigure}
	\hfill
	\begin{subfigure}[b]{0.48\textwidth}
		\centering
		\includegraphics[width=\textwidth]{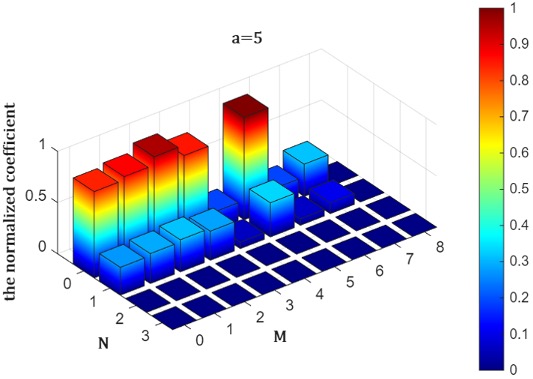}
	\end{subfigure}
	\caption{Distribution of the normalized coefficients ${\left|L_{\left(l\right)}\left(N,M\right)\right|}/{\text{max}\left(\left|L_{\left(l\right)}\left(N,M\right)\right|\right)}$ with respect to $N$ and $M$: $a=0.5$, $a=1$, $a=3$, and $a=5$, when $l=1$.}
	\label{fig4}
\end{figure}

The Legendre-based approach, however, offers a dramatically lighter computational burden. This is evidenced by the Legendre spectrum of $\psi_{\left(l\right)}$, $L_{\left(l\right)}$, obtained from Eq. (\ref{eq3.15}) or Eq. (\ref{eq4.3}). We examine the spectral distribution using the normalized quantity

\begin{align}\label{eq5.1}
\begin{split}
\frac{\left|L_{\left(l\right)}\left(N,M\right)\right|}{\text{max}\left(\left|L_{\left(l\right)}\left(N,M\right)\right|\right)},
\end{split}
\end{align}

\noindent whose decay is shown in figure \ref{fig4} for the first-order convolution ($l=1$) at four representative scattering parameters ($a=0.5,1,3,5$), analogous to a wavenumber spectrum in the horizontal plane. The spectra exhibit two striking features. First, the energy in the vertical Legendre order $N$ decays extremely rapidly: by the third mode ($N=2$) the normalized coefficients have already dropped at ${10}^{-4}$ of the leading term, regardless of $a$. This rapid spectral decay reflects the underlying physics of linear wave diffraction. The vertical structure of the velocity potential on the cylinder surface is governed by the smooth, monotonically decaying (or hyperbolic-cosine) depth dependence imposed by the free-surface conditions. Such a smooth profile is optimally represented by the lowest-order Legendre polynomials, which are themselves smooth and complete on $\left[-1,1\right]$. Consequently, only a handful of vertical modes suffice to capture the entire depth-wise variation with negligible truncation error. Second, the azimuthal spectrum (indexed by $M$) broadens systematically with increasing $a$. For long waves ($a=0.5$) the energy is concentrated in the lowest azimuthal modes ($M=0,1$), consistent with the nearly uniform pressure distribution around the cylinder. As the scattering parameter grows ($a=3$ and $5$), higher-order angular modes become significant because shorter waves excite stronger diffraction and sharper angular gradients—particularly the pronounced contrast between the weather side (high run-up) and the lee-side shadow zone. This multipole-like behavior is exactly what classical scattering theory predicts: higher $a$ requires more angular harmonics to resolve the increasingly oscillatory azimuthal dependence of the scattered potential. The observed increase of the required azimuthal resolution $N_m$ with $a$ is fully consistent with our earlier findings \citep{ShangChenLinWang2026} and further confirms that the Legendre spectral basis correctly captures the geometry- and wavelength-dependent physics of wave–cylinder interaction without any loss of fidelity. The combination of these two behaviors — near-perfect sparsity in the vertical direction and controllable (though $a$-dependent) spread in the azimuthal direction — enables an extraordinarily compact representation. The vertical matrix $\left[D\right]_{N_l\times N_l}$ can be safely truncated to a $3\times3$ system while retaining full accuracy, compared with the $1024\times1024$ matrix required by the Fourier-based method. This reduction (more than five orders of magnitude in matrix size) is the primary source of the Legendre approach’s superior efficiency.

\begin{figure}
	\centering
	\includegraphics[width=0.8\textwidth]{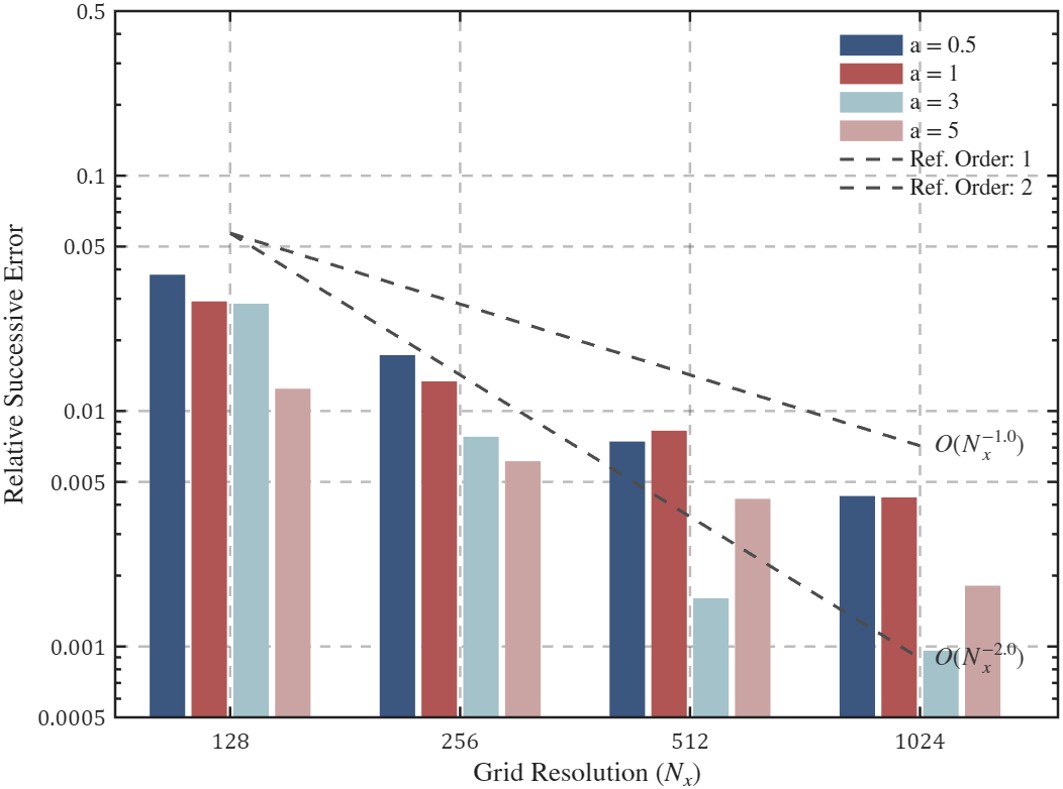}
	\caption{Relative errors of different division nodes $N_{x}$ per wavelength when $l=1$.}
	\label{fig5}
\end{figure}

We investigate the convergence behavior of $\psi_{\left(l\right)}$ as a function of the spatial resolution. The free surface is discretized over a length spanning $N_\lambda$ incident wavelengths, and spatial resolution is controlled by the parameter $N_x$, representing the number of discretization nodes per wavelength. To assess the sensitivity of the solution to the parameter $N_x$, we define a relative successive error functional based on the Frobenius norm:

\begin{align}\label{eq5.2}
\begin{split}
\text{Err}\left(N_{x,i}\right)=\frac{{\parallel\psi_{\left(l\right)}\left(N_{x,i}\right)-\psi_{\left(l\right)}\left(N_{x,i+1}\right)\parallel}_F}{{\parallel\psi_{\left(l\right)}\left(N_{x,i+1}\right)\parallel}_F},
\end{split}
\end{align}

\noindent where $N_{x,i}$ and $N_{x,i+1}$ denote successive levels of refinement. Figure \ref{fig5} illustrates the convergence results for the cases of $a=0.5,1,3,5$.  Similar convergence patterns were observed across all examined wave cases: the numerical uncertainty demonstrates a consistent downward trend as $N_x$ increases from $2^7$ to $2^{10}$ points per wave length. The reference lines labeled "Ref. Order: $O(N_x^{-1.0})$" and "Ref. Order: $O(N_x^{-2.0})$" indicate that the theoretical convergence rate expected for the numerical scheme is between first-order and second-order. Increasing $N_x$ yields a more accurate representation of the cylinder boundary, i.e., the circle, thereby improving the overall calculation accuracy. As shown in figure \ref{fig5}, higher values of $N_x$ enhance the fitting accuracy for all wave cases examined. Specifically, a resolution of $N_x=2^9$ ensures that the relative successive error remains below $1\%$, indicating that the solution has reached a state of grid independence. While further refinement to $N_x=2^{10}$ or $2^{11}$ yields a marginal reduction in error, the associated computational cost increases substantially. Normally, the global spectral differentiation approach yields exponential convergence for globally smooth and periodic fields. However, its accuracy may deteriorate in the immediate vicinity of the solid cylinder boundary ($\rho=a$). The imposed no-penetration boundary condition introduces localized geometric effects associated with the curved solid boundary, which require relatively higher spatial resolution to achieve strict convergence compared with conventional smooth periodic problems. Therefore, the high-resolution cases considered above provide a conservative reference for assessing the numerical convergence behavior. Further investigation shows that, for the present wave–cylinder interaction problems, a much lower resolution of $N_x=64$ points per wavelength can already maintain satisfactory accuracy, with negligible differences compared with the converged solutions. This resolution is therefore adopted in subsequent simulations to balance computational efficiency and numerical accuracy.

\begin{figure}
	\centering
	\begin{subfigure}[b]{\textwidth}
		\centering
		\includegraphics[height=0.17\textheight]{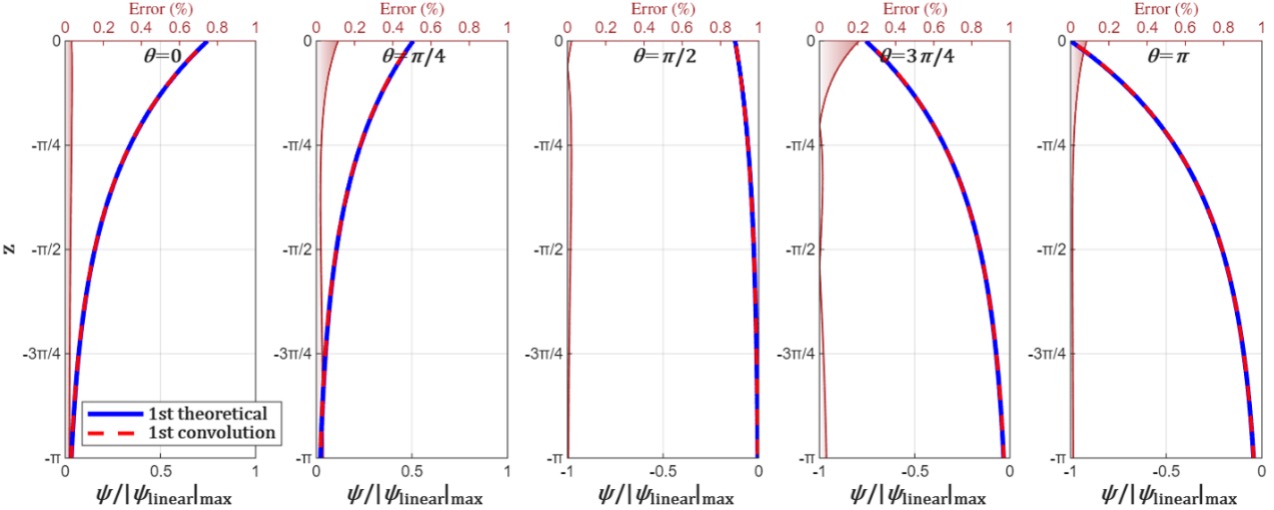}
		\caption{$a=0.5$}
		\label{fig6a}
	\end{subfigure}
	\begin{subfigure}[b]{\textwidth}
		\centering
		\includegraphics[height=0.17\textheight]{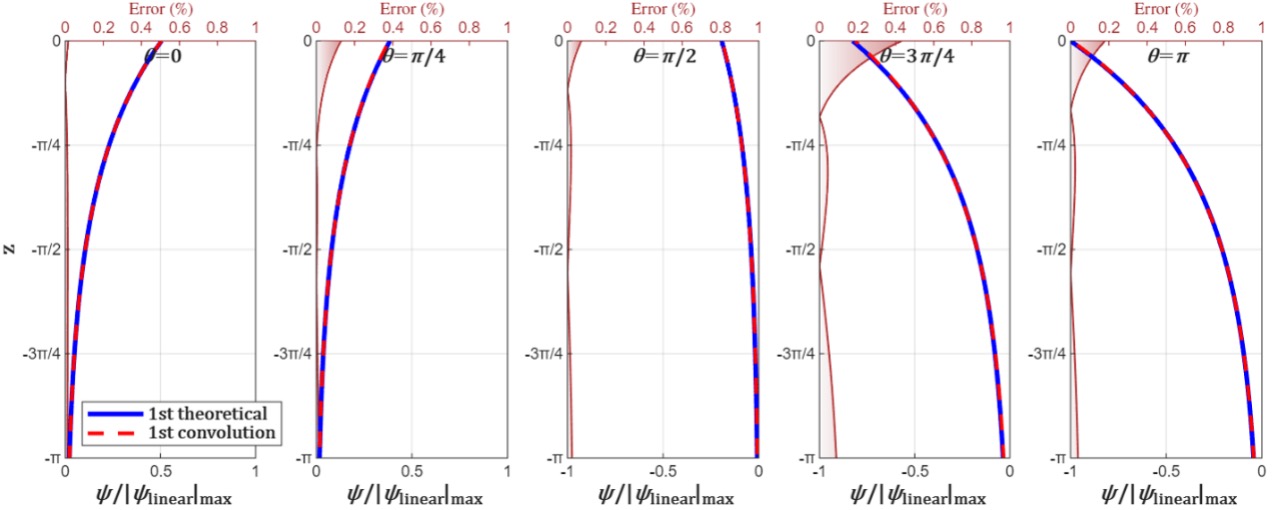}
		\caption{$a=1$}
		\label{fig6b}
	\end{subfigure}
	\begin{subfigure}[b]{\textwidth}
		\centering
		\includegraphics[height=0.17\textheight]{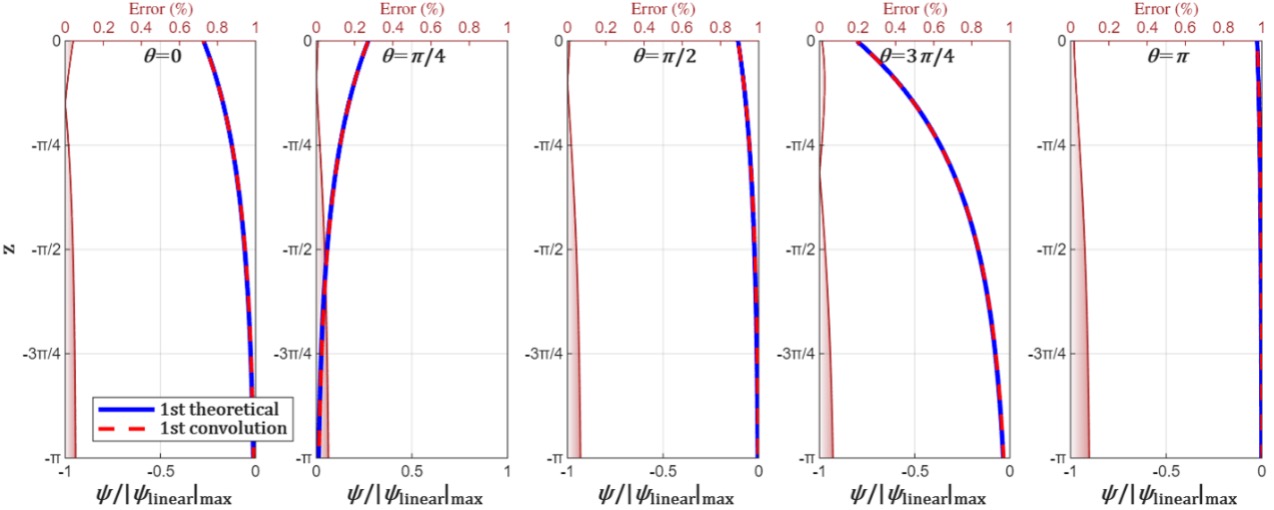}
		\caption{$a=3$}
		\label{fig6c}
	\end{subfigure}
	\begin{subfigure}[b]{\textwidth}
		\centering
		\includegraphics[height=0.17\textheight]{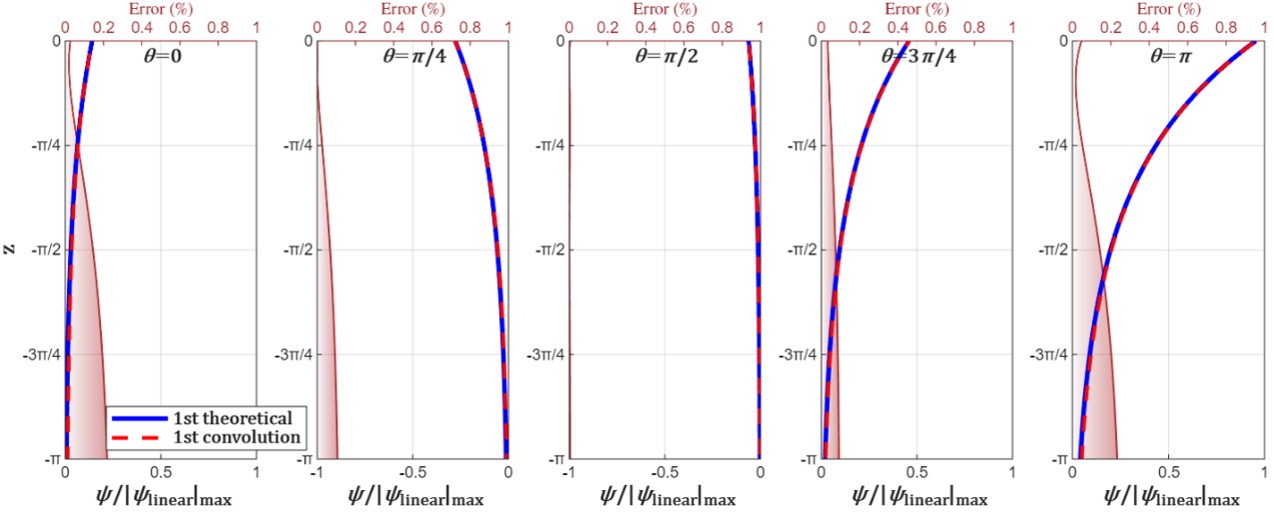}
		\caption{$a=5$}
		\label{fig6d}
	\end{subfigure}
	\caption{Comparison between $\psi_{(1)}$ and linear theoretical solutions for small-steepness incident wave cases.}
	\label{fig6}
\end{figure}

To rigorously verify the proposed methodology, the first-order convolutions of the velocity potential, $\psi_{\left(1\right)}$, are compared against the exact analytical solutions $\psi_{\text{linear}}$ derived from the linear diffraction theory \cite[e.g.][]{Havelock1940}. Figure \ref{fig6} illustrates the vertical profiles of normalized $\psi/\left|\psi_{\text{linear}}\right|_{\text{max}}$ alongside the corresponding global relative error distributions, defined by

\begin{align}\label{eq5.3}
\begin{split}
\text{Err}_1\left(z,\theta\right)=\frac{\left|\psi_{\left(1\right)}\left(z,\theta\right)-\psi_{\text{linear}}\left(z,\theta\right)\right|}{\left|\psi_{\text{linear}}\left(z,\theta\right)\right|_{\text{max}}}\times100\%,
\end{split}
\end{align}

\noindent along the water column ($z\in\left[-\pi,0\right]$). The comparisons are conducted at five discrete azimuthal angles ($\theta=0,\pi/4,\pi/2,3\pi/4,\pi$) and span a broad range of dimensionless scattering parameters $a$ from the long-wave regime ($a=0.5$) to the strongly scattering short-wave regime ($a=5$).  Overall, the numerical results obtained via the first-order convolution (red dashed lines) exhibit an exceptionally high degree of fidelity to the theoretical predictions (blue solid lines). Across all examined scenarios, the curves are virtually indistinguishable. Quantitative analysis reveals that the global relative error remains strictly bounded below $0.5\%$ in every single case. For smaller cylinders ($a=0.5$ and $a=1$), the errors are predominantly confined to the vicinity of the free surface ($z=0$) and decay rapidly with the increasing water depth. For larger cylinders ($a=3$ and $a=5$) the errors are quite small ($<0.1\%$) at $z=0$ and slightly increase towards the deep sea (remaining $<0.3\%$). In the long-wave regime (small $a$), diffraction is weak and the velocity potential is dominated by the incident wave, whose vertical gradients are strongest near the free surface due to the dynamic free-surface boundary condition. Any minor numerical approximations or truncation effects in the depth-dependent convolution kernels therefore manifest most prominently near $z=0$ before decaying rapidly downward. In the short-wave regime (large $a$), strong diffraction occurs, but the potential decays exponentially with depth ($\sim e^{kz}$ in deep water). At greater depths the absolute magnitude of $\psi$ becomes extremely small; even tiny absolute residuals in the convolution integral are therefore amplified in the relative error metric, although these residuals remain negligible in absolute terms and do not affect the overall hydrodynamic fidelity. Strong diffraction conditions typically induce severe numerical instabilities in conventional solvers \cite[e.g., irregular-frequency singularities in boundary-element methods or numerical dispersion/dissipation in CFD, see e.g.,][]{LiangHousseineChenShao2020,MackayLiangJohanning2021}. The present convolution scheme remains robust, maintaining high global accuracy even under these challenging regimes.

From a physical perspective, the proposed model successfully captures the fundamental hydrodynamic features of wave-cylinder interaction. Firstly, the classic exponential attenuation of the velocity potential with increasing water depth is perfectly reproduced, confirming that the depth-dependent integration kernels function as theoretically expected. Secondly, the model accurately resolves the angular distribution of the scattered wave field. The amplitude and phase of $\psi$ vary significantly with the azimuthal angle $\theta$, reflecting the transition from the weather side (wave impact and run-up) to the lee side (shadow zone). The precise overlapping of the curves confirms the method's capability to model these complex, geometry-driven kinematic variations without phase lagging or amplitude dissipation. The excellent agreement in the first-order problem carries significant implications for the overarching modeling framework. By proving that the convolution method can independently and accurately map the underlying hydrodynamics—without relying on computationally expensive volumetric meshing (like BEM or CFD)—this verification lays a solid mathematical and computational foundation. It demonstrates that the spatial memory effects and boundary interactions are properly conserved, thereby justifying the extension of this specific convolution architecture to solve the much more complex second-order or fully nonlinear scattering problems in subsequent sections.

\subsection{Second-order convolutions}

\begin{figure}
	\centering
	\begin{subfigure}[b]{0.48\textwidth}
		\centering
		\includegraphics[width=\textwidth]{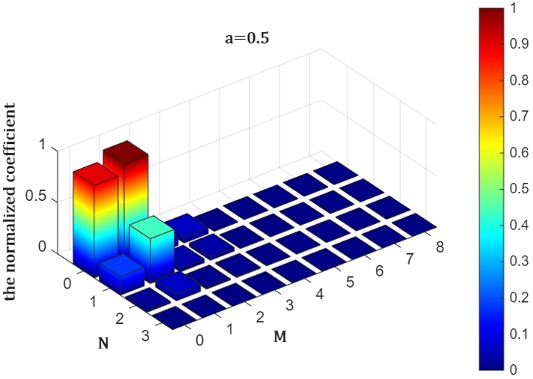}
	\end{subfigure}
	\hfill
	\begin{subfigure}[b]{0.48\textwidth}
		\centering
		\includegraphics[width=\textwidth]{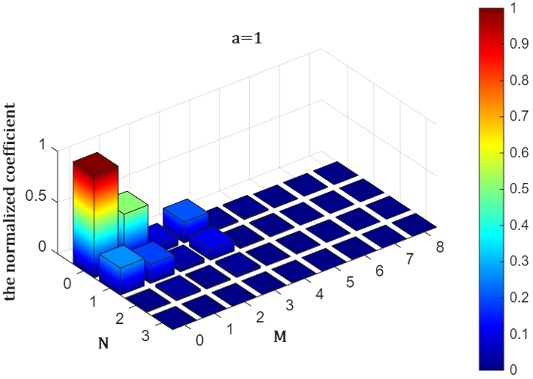}
	\end{subfigure}
	\begin{subfigure}[b]{0.48\textwidth}
		\centering
		\includegraphics[width=\textwidth]{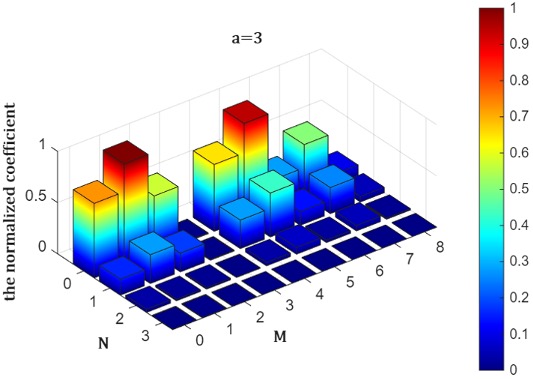}
	\end{subfigure}
	\hfill
	\begin{subfigure}[b]{0.48\textwidth}
		\centering
		\includegraphics[width=\textwidth]{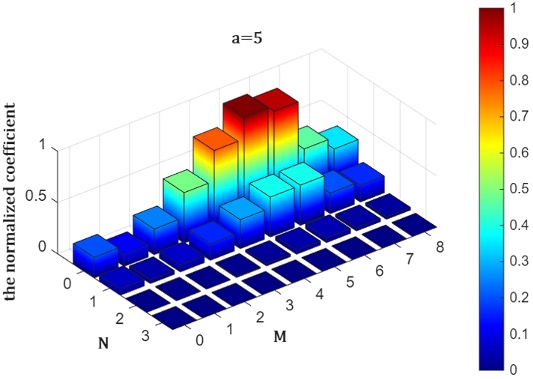}
	\end{subfigure}
	\caption{Distribution of the normalized coefficients ${\left|L_{\left(l\right)}\left(N,M\right)\right|}/{\text{max}\left(\left|L_{\left(l\right)}\left(N,M\right)\right|\right)}$ with respect to $N$ and $M$: $a=0.5$, $a=1$, $a=3$, and $a=5$, when $l=2$.}
	\label{fig7}
\end{figure}

We examine the spectral distribution using the normalized quantity in Eq. (\ref{eq5.2}), whose decay is shown in figure \ref{fig7} for the second-order convolution $l=2$ at four representative scattering parameters ($a=0.5,1,3,5$). The spectra exhibit two striking features. First, the energy in the vertical Legendre order $N$ decays extremely rapidly: by the third mode ($N=2$) the normalized coefficients have already dropped below ${10}^{-4}$ of the leading term, regardless of scattering parameters $a$. This rapid spectral decay reflects the underlying physics of nonlinear wave diffraction. Even for second-order terms, the vertical structure of the velocity potential on the cylinder surface is governed by the smooth, monotonically decaying (or hyperbolic-cosine-like) depth dependence imposed by the quadratic free-surface boundary conditions at double frequency. Such a smooth profile is optimally represented by the lowest-order Legendre polynomials, which are themselves smooth and complete on $\left[-1,1\right]$. Consequently, only a handful of vertical modes suffice to capture the entire depth-wise variation with negligible truncation error. Second, the azimuthal spectrum (indexed by $M$) broadens systematically with increasing $a$. For long waves ($a=0.5$) the energy is concentrated in the lowest azimuthal modes ($M\le1$), consistent with the nearly uniform pressure distribution around the cylinder. As the scattering parameter grows ($a=1,3,5$), second-order angular modes become significant because the nonlinear scattering excites stronger double-frequency components and sharper angular gradients — particularly the pronounced contrast between the weather side (high run-up) and the lee-side shadow zone. This multipole-like behavior is exactly what second-order scattering theory predicts: larger scattering parameters a requires more angular harmonics to resolve the increasingly oscillatory azimuthal dependence of the scattered potential. The observed increase of the required azimuthal resolution with a further confirms that the Legendre spectral basis correctly captures the geometry- and wavelength-dependent physics of nonlinear wave–cylinder interaction without any loss of fidelity. The combination of these two behaviors — near-perfect sparsity in the vertical direction and controllable (though $a$-dependent) spread in the azimuthal direction — enables an extraordinarily compact representation. The vertical matrix can be safely truncated to a small system while retaining full accuracy, compared with the large matrix required by the Fourier-based method. This reduction (more than five orders of magnitude in matrix size) remains the primary source of the Legendre approach’s superior efficiency, now demonstrated for the more demanding second-order convolutions. This spectral analysis for the second-order terms directly parallels the first-order results and confirms that the proposed convolution scheme extends naturally and efficiently to nonlinear problems, laying the foundation for subsequent fully nonlinear scattering simulations.

\begin{figure}
	\centering
	\includegraphics[width=0.8\textwidth]{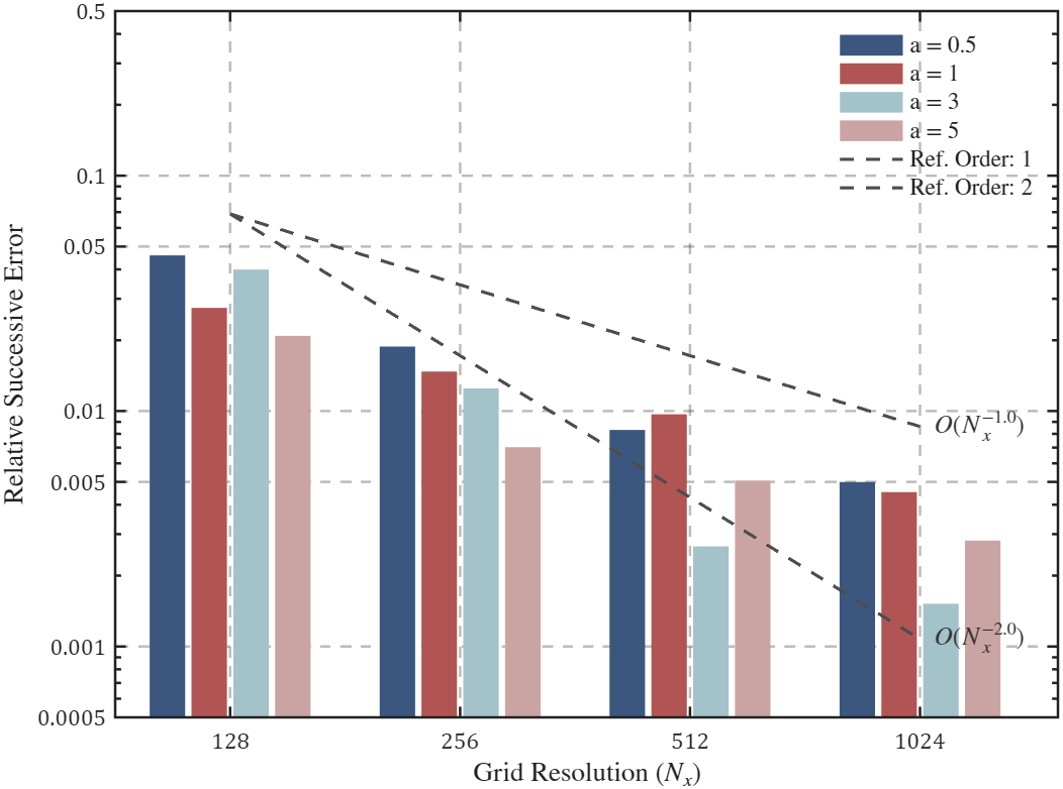}
	\caption{Relative errors of different division nodes $N_{x}$ per wavelength when $l=2$.}
	\label{fig8}
\end{figure}

We then investigate the convergence behavior of the second-order convolution of velocity potential, $\psi_{\left(2\right)}$, as a function of the spatial resolution. The free surface is discretized over a length spanning the same number of incident wavelengths as in the first-order study, and spatial resolution is controlled by the parameter $N_x$, representing the number of discretization nodes per wavelength. To assess the sensitivity of the solution to the parameter $N_x$, the same relative successive error functional in the first-order study is defined based on the Frobenius norm. Figure \ref{fig8} illustrates the convergence results for the second-order convolution ($l=2$). Similar convergence patterns between first- and second- order convergence rate were observed across all examined wave cases: the numerical uncertainty demonstrates a consistent downward trend as $N_x$ increases from $2^7$ to $2^{10}$ points per wave length. Increasing $N_x$ yields a more accurate representation of the cylinder boundary, i.e., the circle, thereby improving the overall calculation accuracy. As shown in figure \ref{fig8}, higher values of $N_x$ enhance the fitting accuracy for all wave cases examined. Specifically, a resolution of $N_x=2^9$ ensures that the relative successive error remains below $1\%$ for all scattering parameters, indicating that the solution has reached a state of grid independence. While further refinement to $N_x=2^{10}$ yields a marginal reduction in error, the associated computational cost increases substantially. Similar to the first-order results, the convergence analysis confirms that the high-resolution cases serve as a conservative reference for verification purposes, whereas the actual recursive computations can be performed with a significantly lower spatial resolution while maintaining sufficient accuracy.

\begin{figure}
	\centering
	\begin{subfigure}[b]{\textwidth}
		\centering
		\includegraphics[height=0.17\textheight]{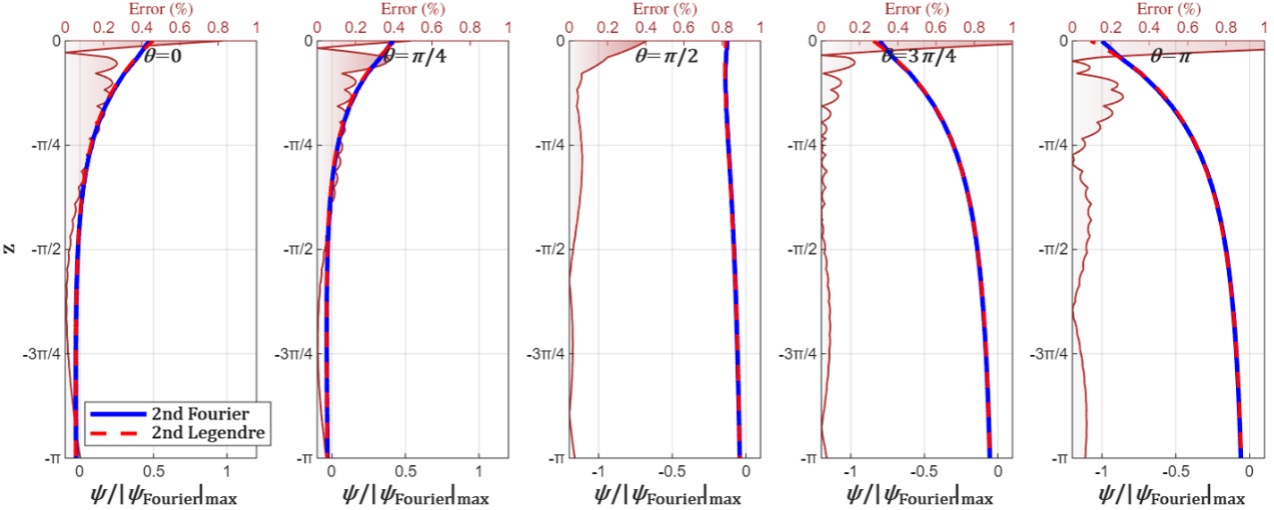}
		\caption{$a=0.5$}
		\label{fig9a}
	\end{subfigure}
	\begin{subfigure}[b]{\textwidth}
		\centering
		\includegraphics[height=0.17\textheight]{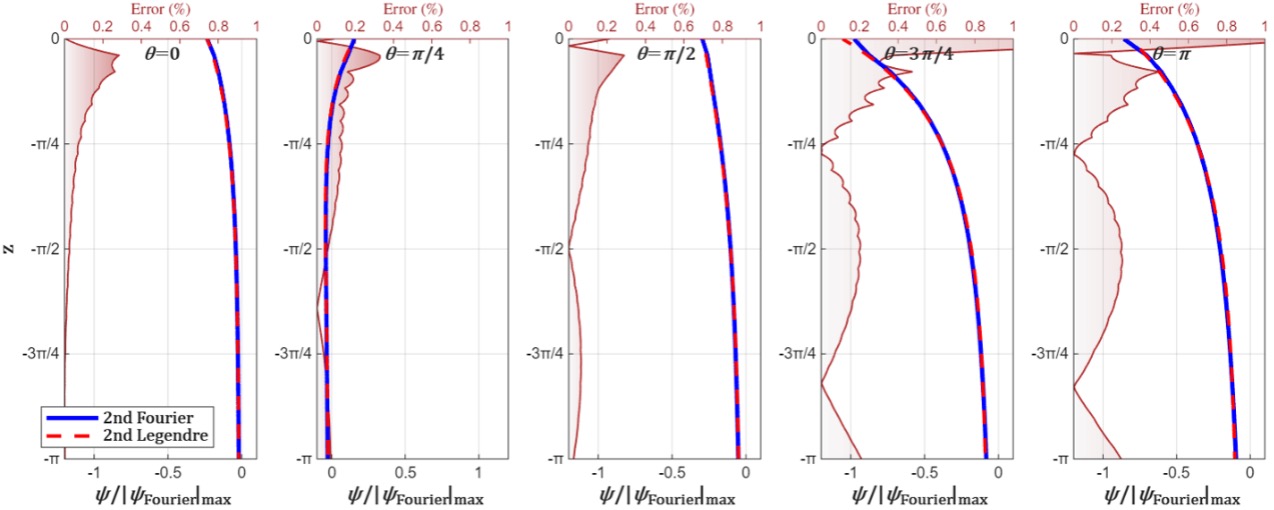}
		\caption{$a=1$}
		\label{fig9b}
	\end{subfigure}
	\begin{subfigure}[b]{\textwidth}
		\centering
		\includegraphics[height=0.17\textheight]{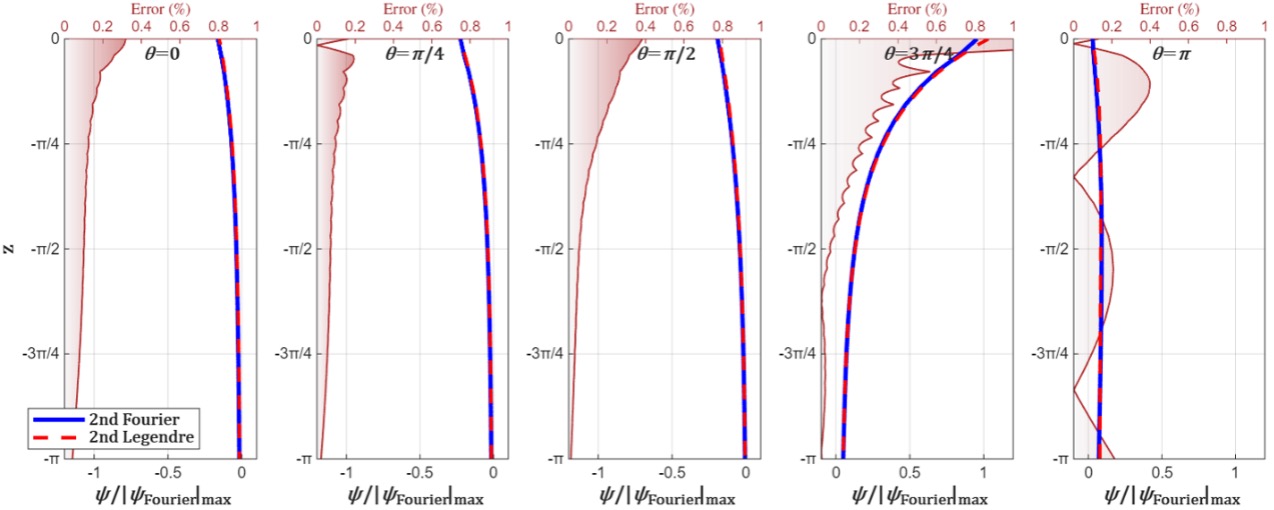}
		\caption{$a=3$}
		\label{fig9c}
	\end{subfigure}
	\begin{subfigure}[b]{\textwidth}
		\centering
		\includegraphics[height=0.17\textheight]{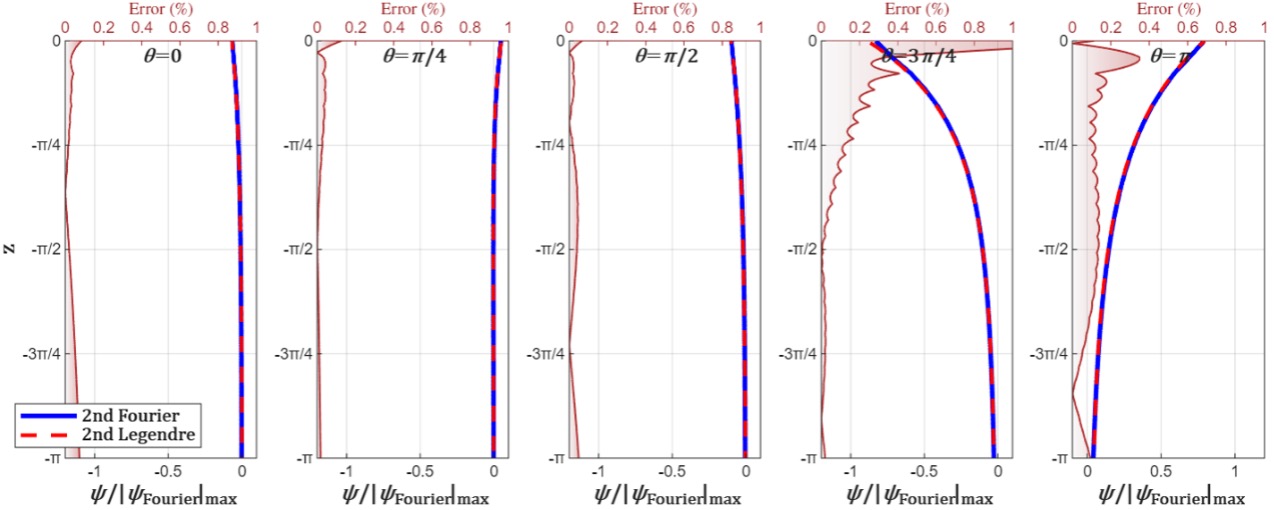}
		\caption{$a=5$}
		\label{fig9d}
	\end{subfigure}
	\caption{Comparison of $\psi_{(2)}$ between the Fourier- and the Legendre-based solutions for small-steepness incident wave cases.}
	\label{fig9}
\end{figure}

To rigorously verify the proposed methodology for nonlinear problems, the second-order convolutions of the velocity potential, $\psi_{\left(2\right)}$, obtained via the Legendre-based formulation are compared directly against the corresponding Fourier-based solutions. The comparison is performed between the two variants of the novel scheme itself since existing second-order theories rely on Stokes perturbation expansions rather than the present convolution architecture. To be noted, the convolution-based expansion leads to decomposition that converge essentially different from perturbation approach, which has been comprehensively discussed in \cite{Wang2025}. Figure \ref{fig9} illustrates the vertical profiles of normalized $\psi_{\left(2\right)}$ at five discrete azimuthal angles ($\theta=0,\pi/4,\pi/2,3\pi/4,\pi$) together with the corresponding global relative error distributions, defined by

\begin{align}\label{eq5.4}
\begin{split}
\text{Err}_2\left(z,\theta\right)=\frac{\left|\psi_{\text{Legendre}}\left(z,\theta\right)-\psi_{\text{Fourier}}\left(z,\theta\right)\right|}{\left|\psi_{\text{Fourier}}\left(z,\theta\right)\right|_{\text{max}}}\times100\%,
\end{split}
\end{align}

\noindent along the water column ($z\in[-\pi,0]$). The comparisons span the same broad range of dimensionless scattering parameters a from the long-wave regime ($a=0.5$) to the strongly scattering short-wave regime ($a=5$).

Overall, the Fourier- and Legendre-based solutions exhibit excellent agreement across all examined cases, with the curves virtually overlapping for the majority of the water column. Quantitative analysis reveals that the global relative error remains most of the regions bounded below $1\%$ in every single case. For either smaller or larger cylinders the discrepancies are predominantly confined to the immediate vicinity of the free surface ($z\approx0$) and decay rapidly with increasing water depth; The errors are small ($<0.5\%$) as long as departing from the free surface. The discrepancies near the free surface are readily explained by the differing representation properties of the two vertical bases. The second-order potential satisfies quadratic free-surface boundary conditions that impose strong vertical gradients and an essentially exponential ($e^{kz}$-like) depth dependence in the near-surface layer. The Fourier-based vertical expansion, being global and oscillatory, requires a significantly larger number of modes to resolve these sharp near-surface features without Gibbs-type oscillations or truncation error. In contrast, the Legendre polynomials, being complete and optimally suited to the smooth yet rapidly varying profile on the stretched coordinate $\left[-1,1\right]$, capture the same physics with far fewer modes. Consequently, the Legendre-based solution converges more rapidly near $z=0$, while the Fourier-based solution shows a residual mismatch unless a prohibitively large number of vertical modes is retained. Away from the free surface, where the potential decays smoothly, both formulations yield identical results.

From a physical perspective, the proposed convolution scheme successfully captures the fundamental hydrodynamic features of second-order wave-cylinder interaction. Firstly, the double-frequency depth dependence and exponential attenuation of $\psi_{\left(2\right)}$ with increasing water depth are faithfully reproduced by both approaches, confirming that the depth-dependent convolution kernels operate as theoretically expected even under nonlinear conditions. Secondly, the model accurately resolves the angular distribution of the scattered double-frequency field. The amplitude and phase of $\psi_{\left(2\right)}$ vary markedly with the azimuthal angle $\theta$, reflecting the transition from the weather side (enhanced nonlinear run-up) to the lee-side shadow zone. The close overlap of the two independent solutions confirms the method’s robustness in modeling these complex, geometry-driven kinematic variations without phase lagging or amplitude dissipation. The excellent consistency between the Fourier- and Legendre-based formulations for the second-order problem carries significant implications for the overarching modeling framework. By demonstrating that the convolution architecture can be extended reliably to nonlinear scattering — without relying on computationally expensive volumetric meshing (like BEM or CFD) — this verification further strengthens the mathematical and computational foundation established in the first-order study. It shows that spatial memory effects and boundary interactions are conserved at second order. The proposed framework can be further extended to higher-order nonlinearities (including third- and higher-order terms) by iteratively updating the corresponding coefficients $\xi_{\left(l\right)}$ (see Eq. (\ref{eq2.39})). In the present study, the verification is limited to the second-order contribution, which provides the essential validation of the nonlinear convolution framework. The accuracy assessment and physical analysis of higher-order terms will be investigated separately in Part II.

\section{Conclusion}\label{sec6}

This paper has presented a novel and efficient fully nonlinear time-domain model for simulating wave diffraction by a vertical cylinder, grounded in potential flow theory and the Enhanced Spectral Boundary Integral (ESBI) method. The primary contribution is the development of a robust solver for the cylinder surface velocity potential, which is the key to accurately computing wave loads and free surface kinematics. From the theoretical development, numerical implementation, and verification studies, the following conclusions can be drawn:

a) A unified spectral framework: The boundary value problem for the cylinder surface potential $\psi$ was successfully formulated as a second-kind Fredholm integral equation (Eq. \ref{eq2.38}). Two spectral approaches were developed to solve this equation: a Fourier mode-based method (Eq. \ref{eq3.7}) and a Legendre polynomial-based method (Eq. \ref{eq3.15}), both of which transform the problem into solving a system of linear equations in spectral space.

b) Verification and accuracy: The model was rigorously verified against first-order linear diffraction theory \citep{Havelock1940} for small-steepness incident waves. The computed velocity potential $\psi$ showed excellent agreement with theoretical predictions for various wave conditions and across all azimuthal angles, confirming the fundamental accuracy of the proposed solvers. The analysis confirmed that computation of second-order convolution terms are also robust in this regime, justified by the two different numerical modes.

c) Exceptional efficiency of the Legendre-based method: A key finding of this work is the dramatically superior computational efficiency of the Legendre-based approach. Analysis of the Legendre spectrum $L_{\left(l\right)}$ revealed an extremely rapid decay of the normalized coefficients, reaching the order of $2^{-4}$ of the leading term by the third Legendre mode ($N=2$). This remarkable property permits the truncation of the matrix $\left[D\right]_{N_l\times N_l}$ to a mere $3\times3$ system, yielding a profound decrease in computational cost and memory requirements.

d) Convergence and discretization strategies: A thorough convergence analysis was performed. It was established that for accurate matrix computation, the integral over the wavenumber $k$ must be treated carefully. By setting a truncation limit of $k_t=5\times{10}^4$ and using a discretization step of $\Delta k={10}^{-2}$, the integrals converge with errors below ${10}^{-4}$. 

In summary, this work has developed and verified a new efficient numerical framework for solving nonlinear wave diffraction problems due to vertical circular cylinders. The proposed Legendre-based formulations provide accurate and computationally efficient solutions for the first- and second-order boundary-value problems, establishing a promising approach for further investigations of fully nonlinear wave scattering in directional wave field accounting for surface-piercing structures in a pseudo-spectrum based model.

The present study focuses primarily on the verification of the proposed boundary-value solver under baseline hydrodynamic conditions, demonstrating its precision and numerical stability. Building upon this verified framework, the Part II of this work will extend the investigation to a broader range of physical parameters, including higher wave steepness and various wave scattering regimes. Furthermore, the model will be applied to transient wave diffraction problems in three-dimensional configurations, exploring the spatio-temporal nonlinear wave dynamics across diverse wave conditions and structural scales.

\appendix
\section{Formulations of $I_S$ and $I_C$}\label{appA}
The formula for $I_S$ is given as:

\begin{align}\label{a1}
\begin{split}
\mathcal{F}\left\{I_S\right\}=\sum_{u=1}^{\mathcal{M}} \frac{(-1)^{u-1}}{(u-1)!} k \mathcal{F}\left\{\eta^{u-1} \mathcal{F}^{-1}\left\{k^{u-1}(\Phi-\Psi)\right\}\right\},
\end{split}
\end{align}

\noindent where

\begin{align}\label{a2}
\begin{split}
\Phi(\boldsymbol{k}) =-\frac{\partial}{\partial K} \iint_C \varphi^{\prime} e^{-\text{i} K \cos \left(\alpha-\theta^{\prime}\right)+k y^{\prime}} \text{d} s^{\prime}.
\end{split}
\end{align}

The formula for $I_C$ is given as:

\begin{align}\label{a3}
\begin{split}
& I_C-\mathcal{F}_C^{-1}\left\{\frac{\partial}{\partial K} \int_0^{2 \pi}\right.  \left.\int_{-\infty}^0 I_C^{\prime} e^{-\text{i} K \cos \left(\alpha-\theta^{\prime}\right)} e^{-k\left|z^{\prime}-z\right|} \text{d} z^{\prime} a \text{d} \theta^{\prime}\right\} \\ 
& =\mathcal{F}_C^{-1}\left\{\left(e^{k y}-e^{k z}\right)\left[\sum_{u=1}^{\mathcal{M}} \frac{(-1)^{u-1}}{(u-1)!}\left[k^{u-2} \mathcal{F}\left\{\eta^{u-1} V\right\}-k^{u-3} \mathrm{i} \boldsymbol{k} \cdot \mathcal{F}\{\nabla \phi\}\right]+\sum_{u=1}^{\mathcal{M}-1} \frac{(-1)^{u-1}}{u!} k^{u-1} \chi_u\right]\right\}\\ 
& +\mathcal{F}_C^{-1}\left\{\frac{\partial}{\partial K}\iint_C \varphi^{\prime} e^{-\text{i} K \cos \left(\alpha-\theta^{\prime}\right)}e^{-k\left|y^{\prime}-y\right|} \text{d} s^{\prime}\right.  \left.-\frac{\partial}{\partial K} \int_0^{2 \pi} \int_{-\infty}^0 \psi^{\prime} e^{-\text{i} K \cos \left(\alpha-\theta^{\prime}\right)} e^{-k\left|z^{\prime}-z\right|} \text{d} z^{\prime} a \text{d} \theta^{\prime}\right\}.
\end{split}
\end{align}

\section{Formulations of $\Im_{(l)}$}\label{appC}
The formulations for $\Im_{(l)}$ at each order $l>1$ are obtained as:

\begin{align}\label{c1}
\begin{split}
\mathfrak{J}_{(l)}=\mathcal{L}_{(l)}+\sum_{u=1}^{l-2} \frac{(-1)^{u-1}}{u!} \mathcal{F}\left\{\eta^u \mathcal{F}^{-1}\left\{k^u \Im_{(l-u)}\right\}\right\},
\end{split}
\end{align}

\noindent where

\begin{align}\label{c2}
\begin{split}
\mathcal{L}_{(l)}=-\sum_{j=1}^{(\mathcal{M}-1)/2} \sum_{u=1}^{2 j} \frac{(-1)^{u-1}\langle l > (2 j-1)\rangle}{(u-1)!(2 j-u)!}  \mathcal{F}\left\{\eta^{u-1} \mathcal{F}^{-1}\left\{k^{2 j-2} \chi_{2 j-u(l-2 j+1)}\right\}\right\}.
\end{split}
\end{align}

\noindent $\langle l > (2 j-1)\rangle$ is equal to 1 if $l$ is larger than $(2 j-1)$, else it is equal to 0.

\bibliographystyle{elsarticle-num}
\bibliography{cas-refs}

@article{AggarwalTomaselliChristensenBihs2020,
  author  = {Aggarwal, A. and Tomaselli, P. D. and Christensen, E. D. and Bihs, H.},
  title   = {Computational fluid dynamics investigations of breaking focused wave-induced loads on a monopile and the effect of breaker location},
  journal = {Journal of Offshore Mechanics and Arctic Engineering},
  volume  = {142},
  pages   = {021903},
  year    = {2020}
}

@inproceedings{AliyarDucrozetBouscasseVenkatachalamFerrant2022,
  author    = {Aliyar, S. and Ducrozet, G. and Bouscasse, B. and Venkatachalam, S. and Ferrant, P.},
  title     = {Breaking focused wave interaction with cylinder using HOS-OpenFOAM coupling},
  booktitle = {OCEANS 2022-Chennai},
  publisher = {IEEE},
  pages     = {1--10},
  year      = {2022}
}

@article{BaiEatockTaylor2007,
  author  = {Bai, W. and Eatock-Taylor, R.},
  title   = {Numerical simulation of fully nonlinear regular and focused wave diffraction around a vertical cylinder using domain decomposition},
  journal = {Applied Ocean Research},
  volume  = {29},
  pages   = {55--71},
  year    = {2007}
}

@article{BaiTeng2013,
  author  = {Bai, W. and Teng, B.},
  title   = {Simulation of second-order wave interaction with fixed and floating structures in time domain},
  journal = {Ocean Engineering},
  volume  = {74},
  pages   = {168--177},
  year    = {2013}
}

@article{BihsChellaKamathArntsen2017,
  author  = {Bihs, H. and Chella, M. A. and Kamath, A. and Arntsen, {\O}. A.},
  title   = {Numerical investigation of focused waves and their interaction with a vertical cylinder using REEF3D},
  journal = {Journal of Offshore Mechanics and Arctic Engineering},
  volume  = {139},
  pages   = {041101},
  year    = {2017}
}

@book{Chakrabarti1987,
  author    = {Chakrabarti, S. K.},
  title     = {Hydrodynamics of Offshore Structures},
  publisher = {WIT Press},
  year      = {1987}
}

@article{ChaplinRaineyYemm1997,
  author  = {Chaplin, J. R. and Rainey, R. C. T. and Yemm, R.},
  title   = {Ringing of a vertical cylinder in waves},
  journal = {Journal of Fluid Mechanics},
  volume  = {350},
  pages   = {119--147},
  year    = {1997}
}

@article{ChatjigeorgiouKorobkinCooker2016,
  author  = {Chatjigeorgiou, I. K. and Korobkin, A. A. and Cooker, M. J.},
  title   = {Three-dimensional steep wave impact on a vertical cylinder},
  journal = {Journal of Hydrodynamics},
  volume  = {28},
  pages   = {523--533},
  year    = {2016}
}

@article{ChauEatockTaylor1992,
  author  = {Chau, F. and Eatock-Taylor, R.},
  title   = {Second-order wave diffraction by a vertical cylinder},
  journal = {Journal of Fluid Mechanics},
  volume  = {240},
  pages   = {571--599},
  year    = {1992}
}

@article{ChenDuffour2018,
  author  = {Chen, C. and Duffour, P.},
  title   = {Modelling damping sources in monopile-supported offshore wind turbines},
  journal = {Wind Energy},
  volume  = {21},
  pages   = {1121--1140},
  year    = {2018}
}

@article{ChenZang2014,
  author  = {Chen, L. and Zang, J. and Hillis, A. J. and Morgan, G. C. and Plummer, A. R.},
  title   = {Numerical investigation of wave--structure interaction using OpenFOAM},
  journal = {Ocean Engineering},
  volume  = {88},
  pages   = {91--109},
  year    = {2014}
}

@article{ChenZangTaylorSunMorganGriceOrszaghovaRuiz2018,
  author  = {Chen, L. and Zang, J. and Taylor, P. H. and Sun, L. and Morgan, G. and Grice, J. and Orszaghova, J. and Ruiz, M. T.},
  title   = {An experimental decomposition of nonlinear forces on a surface-piercing column: Stokes-type expansions of the force harmonics},
  journal = {Journal of Fluid Mechanics},
  volume  = {848},
  pages   = {42--77},
  year    = {2018}
}

@article{ChenMaDongMaTaiNiu2020,
  author  = {Chen, Q.-b. and Ma, Y.-x. and Dong, G.-h. and Ma, X.-z. and Tai, B. and Niu, X.-y.},
  title   = {A numerical investigation of interactions between extreme waves and a vertical cylinder},
  journal = {Journal of Hydrodynamics},
  volume  = {32},
  pages   = {853--864},
  year    = {2020}
}

@article{ChildVenugopal2010,
  author  = {Child, B. and Venugopal, V.},
  title   = {Optimal configurations of wave energy device arrays},
  journal = {Ocean Engineering},
  volume  = {37},
  pages   = {1402--1417},
  year    = {2010}
}

@article{DommermuthYue1987,
  author  = {Dommermuth, D. G. and Yue, D. K.},
  title   = {A high-order spectral method for the study of nonlinear gravity waves},
  journal = {Journal of Fluid Mechanics},
  volume  = {184},
  pages   = {267--288},
  year    = {1987}
}

@article{DuanChenZhao2015,
  author  = {Duan, W. and Chen, J. and Zhao, B.},
  title   = {Second-order Taylor Expansion Boundary Element Method for the second-order wave diffraction problem},
  journal = {Engineering Analysis with Boundary Elements},
  volume  = {58},
  pages   = {140--150},
  year    = {2015}
}

@techreport{EngsigKarupMonteserinEskilsson2017,
  author      = {Engsig-Karup, A. and Monteserin, C. and Eskilsson, C.},
  title       = {A Stabilised Nodal Spectral Element Method for Fully Nonlinear Water Waves, Part 2: Wave-body interaction},
  institution = {arXiv},
  number      = {arXiv:1703.09697},
  year        = {2017}
}

@article{FaltinsenNewmanVinje1995,
  author  = {Faltinsen, O. M. and Newman, J. N. and Vinje, T.},
  title   = {Nonlinear wave loads on a slender vertical cylinder},
  journal = {Journal of Fluid Mechanics},
  volume  = {289},
  pages   = {179--198},
  year    = {1995}
}

@inproceedings{Ferrant1998,
  author    = {Ferrant, P.},
  title     = {Fully nonlinear interactions of long-crested wave packets with a three-dimensional body},
  booktitle = {Proc. 22nd ONR Symposium on Naval Hydrodynamics},
  year      = {1998}
}

@article{FitzgeraldTaylorTaylorGriceZang2014,
  author  = {Fitzgerald, C. J. and Taylor, P. H. and Taylor, R. E. and Grice, J. and Zang, J.},
  title   = {Phase manipulation and the harmonic components of ringing forces on a surface-piercing column},
  journal = {Proceedings of the Royal Society A: Mathematical, Physical and Engineering Sciences},
  volume  = {470},
  year    = {2014}
}

@article{GaoZhangZhang2024,
  author  = {Gao, N. and Zhang, Y. and Zhang, H.},
  title   = {CFD Investigations of Interaction Between Freak Wave and Bottom-Mounted Vertical Cylinder},
  journal = {Journal of Shanghai Jiaotong University (Science)},
  volume  = {29},
  pages   = {809--816},
  year    = {2024}
}

@article{GhadirianBredmose2019,
  author  = {Ghadirian, A. and Bredmose, H.},
  title   = {Investigation of the effect of the bed slope on extreme waves using first order reliability method},
  journal = {Marine Structures},
  volume  = {67},
  pages   = {102627},
  year    = {2019}
}

@article{GhadirianBredmoseDixen2016,
  author  = {Ghadirian, A. and Bredmose, H. and Dixen, M.},
  title   = {Breaking phase focused wave group loads on offshore wind turbine monopiles},
  journal = {Journal of Physics: Conference Series},
  volume  = {753},
  pages   = {092004},
  year    = {2016}
}

@article{GhadirianPierellaBredmose2023,
  author  = {Ghadirian, A. and Pierella, F. and Bredmose, H.},
  title   = {Calculation of slamming wave loads on monopiles using fully nonlinear kinematics and a pressure impulse model},
  journal = {Coastal Engineering},
  volume  = {179},
  pages   = {104219},
  year    = {2023}
}

@article{Goda1966,
  author  = {Goda, Y.},
  title   = {A study on impulsive breaking wave force upon a vertical pile},
  journal = {Rept. Port and Harbour Res. Inst.},
  volume  = {5},
  pages   = {1--30},
  year    = {1966}
}

@techreport{GrueBjorsholStrand1993,
  author      = {Grue, J. and Bj{\o}rshol, G. and Strand, {\O}.},
  title       = {Higher harmonic wave exciting forces on a vertical cylinder},
  institution = {Inst. Math., Univ. Oslo},
  year        = {1993}
}

@article{GrueHuseby2002,
  author  = {Grue, J. and Huseby, M.},
  title   = {Higher-harmonic wave forces and ringing of vertical cylinders},
  journal = {Applied Ocean Research},
  volume  = {24},
  pages   = {203--214},
  year    = {2002}
}

@incollection{Grue2008,
  author    = {Grue, J.},
  title     = {Nonlinear Surface Waves interacting with a Vertical Cylinder. The first few Steps of a General Method},
  booktitle = {IUTAM Symposium on Fluid-Structure Interaction in Ocean Engineering},
  publisher = {Springer},
  pages     = {79--92},
  year      = {2008}
}

@article{Havelock1940,
  author  = {Havelock, T. H.},
  title   = {The pressure of water waves upon a fixed obstacle},
  journal = {Proceedings of the Royal Society of London. Series A. Mathematical and Physical Sciences},
  volume  = {175},
  pages   = {409--421},
  year    = {1940}
}

@article{HayatdavoodiNeillErtekin2018,
  author  = {Hayatdavoodi, M. and Neill, D. R. and Ertekin, R. C.},
  title   = {Diffraction of cnoidal waves by vertical cylinders in shallow water},
  journal = {Theoretical and Computational Fluid Dynamics},
  volume  = {32},
  pages   = {561--591},
  year    = {2018}
}

@article{HayatdavoodiErtekin2023,
  author  = {Hayatdavoodi, M. and Ertekin, R. C.},
  title   = {Diffraction and refraction of nonlinear waves by the Green--Naghdi equations},
  journal = {Journal of Offshore Mechanics and Arctic Engineering},
  volume  = {145},
  pages   = {021201},
  year    = {2023}
}

@article{HeWangHeLiFengChabchoub2025,
  author  = {He, Y. and Wang, J. and He, J. and Li, Y. and Feng, X. and Chabchoub, A.},
  title   = {Lifetime characterisation of extreme wave localisations in crossing seas},
  journal = {Journal of Fluid Mechanics},
  volume  = {1008},
  pages   = {A3},
  year    = {2025}
}

@article{HlopheAdcockDingZangDaiTangTaylor2025,
  author  = {Hlophe, T. and Adcock, T. A. and Ding, H. and Zang, J. and Dai, S. and Tang, T. and Taylor, P. H.},
  title   = {Nonlinear wave loads on monopile foundations and structural response in severe wave conditions},
  journal = {Applied Ocean Research},
  volume  = {164},
  pages   = {104790},
  year    = {2025}
}

@article{IsaacsonCheung1992,
  author  = {Isaacson, M. and Cheung, K. F.},
  title   = {Time-domain second-order wave diffraction in three dimensions},
  journal = {Journal of Waterway, Port, Coastal, and Ocean Engineering},
  volume  = {118},
  pages   = {496--516},
  year    = {1992}
}

@article{JiangelMoctarZhang2024,
  author  = {Jiang, C. and {el Moctar}, O. and Zhang, G.},
  title   = {Nonlinear modeling of wave--structure interaction for a flexible floating structure},
  journal = {Ocean Engineering},
  volume  = {300},
  pages   = {117489},
  year    = {2024}
}

@article{JohannessenSwan2001,
  author  = {Johannessen, T. B. and Swan, C.},
  title   = {A laboratory study of the focusing of transient and directionally spread surface water waves},
  journal = {Proceedings of the Royal Society of London. Series A: Mathematical, Physical and Engineering Sciences},
  volume  = {457},
  pages   = {971--1006},
  year    = {2001}
}

@article{JohannessenSwan2003,
  author  = {Johannessen, T. B. and Swan, C.},
  title   = {On the nonlinear dynamics of wave groups produced by the focusing of surface-water waves},
  journal = {Proceedings of the Royal Society of London. Series A: Mathematical, Physical and Engineering Sciences},
  volume  = {459},
  pages   = {1021--1052},
  year    = {2003}
}

@article{KamathChellaBihsArntsen2016,
  author  = {Kamath, A. and Chella, M. A. and Bihs, H. and Arntsen, {\O}. A.},
  title   = {Breaking wave interaction with a vertical cylinder and the effect of breaker location},
  journal = {Ocean Engineering},
  volume  = {128},
  pages   = {105--115},
  year    = {2016}
}

@article{KangZhangYanJuZhuGu2024,
  author  = {Kang, A. and Zhang, D. and Yan, B. and Ju, J. and Zhu, B. and Gu, Y.},
  title   = {Experimental and numerical investigation of focused wave interactions with a vertical truncated dumbbell-shaped cylinder},
  journal = {Ocean Engineering},
  volume  = {292},
  pages   = {116334},
  year    = {2024}
}

@article{KangZengCuiChen2024,
  author  = {Kang, Y. and Zeng, X. and Cui, Z. and Chen, J.},
  title   = {Analytical solutions for hydrodynamic responses of arrays of floating truncated cylinders using multi-term Galerkin method and its application to a new wave energy converter device},
  journal = {Physics of Fluids},
  volume  = {36},
  year    = {2024}
}

@techreport{vonKarman1929,
  author      = {{von Karman}, T.},
  title       = {The impact on seaplane floats during landing},
  institution = {NACA},
  number      = {Technical Note 321},
  year        = {1929}
}

@article{KeuleganCarpenter1958,
  author  = {Keulegan, G. H. and Carpenter, L. H.},
  title   = {Forces on cylinders and plates in an oscillatory flow},
  journal = {Journal of Research of the National Bureau of Standards},
  volume  = {60},
  pages   = {426--440},
  year    = {1958}
}

@article{KimYue1989,
  author  = {Kim, M.-H. and Yue, D. K.},
  title   = {The complete second-order diffraction solution for an axisymmetric body. Part 1. Monochromatic incident waves},
  journal = {Journal of Fluid Mechanics},
  volume  = {200},
  pages   = {235--264},
  year    = {1989}
}

@article{KoivistoGeaBermudezSorensen2020,
  author  = {Koivisto, M. and Gea-Berm{\'u}dez, J. and S{\o}rensen, P.},
  title   = {North Sea offshore grid development: combined optimisation of grid and generation investments towards 2050},
  journal = {IET Renewable Power Generation},
  volume  = {14},
  pages   = {1259--1267},
  year    = {2020}
}

@article{Kriebel1990,
  author  = {Kriebel, D. L.},
  title   = {Nonlinear wave interaction with a vertical circular cylinder. Part I: Diffraction theory},
  journal = {Ocean Engineering},
  volume  = {17},
  pages   = {345--377},
  year    = {1990}
}

@article{Kriebel1992,
  author  = {Kriebel, D. L.},
  title   = {Nonlinear wave interaction with a vertical circular cylinder. Part II: Wave run-up},
  journal = {Ocean Engineering},
  volume  = {19},
  pages   = {75--99},
  year    = {1992}
}

@inproceedings{KristiansenBaarholmStansberg2004,
  author    = {Kristiansen, T. and Baarholm, R. and Stansberg, C. T.},
  title     = {Validation of second-order analysis in predicting diffracted wave elevation around a vertical circular cylinder},
  booktitle = {ISOPE International Ocean and Polar Engineering Conference},
  publisher = {ISOPE},
  year      = {2004}
}

@article{KristiansenFaltinsen2017,
  author  = {Kristiansen, T. and Faltinsen, O. M.},
  title   = {Higher harmonic wave loads on a vertical cylinder in finite water depth},
  journal = {Journal of Fluid Mechanics},
  volume  = {833},
  pages   = {773--805},
  year    = {2017}
}

@article{LiLiu2019,
  author  = {Li, A.-j. and Liu, Y.},
  title   = {New analytical solutions to water wave diffraction by vertical truncated cylinders},
  journal = {International Journal of Naval Architecture and Ocean Engineering},
  volume  = {11},
  pages   = {952--969},
  year    = {2019}
}

@article{LiWangLiu2012,
  author  = {Li, J. and Wang, Z. and Liu, S.},
  title   = {Experimental study of interactions between multi-directional focused wave and vertical circular cylinder, Part I: Wave run-up},
  journal = {Coastal Engineering},
  volume  = {64},
  pages   = {151--160},
  year    = {2012}
}

@article{LiWangLiu2014a,
  author  = {Li, J. and Wang, Z. and Liu, S.},
  title   = {Experimental study of interactions between multi-directional focused wave and vertical circular cylinder, part II: Wave force},
  journal = {Coastal Engineering},
  volume  = {83},
  pages   = {233--242},
  year    = {2014}
}

@article{LiZhangLiuFanZang2022,
  author  = {Li, J. and Zhang, H. and Liu, S. and Fan, Y. and Zang, J.},
  title   = {Experimental investigations of secondary load cycle formation in wave force on a circular cylinder under steep regular waves},
  journal = {Ocean Engineering},
  volume  = {253},
  pages   = {111265},
  year    = {2022}
}

@article{LiangHousseineChenShao2020,
  author  = {Liang, H. and Housseine, C. O. and Chen, X. and Shao, Y.},
  title   = {Efficient methods free of irregular frequencies in wave and solid/porous structure interactions},
  journal = {Journal of Fluids and Structures},
  volume  = {98},
  pages   = {103130},
  year    = {2022}
}

@book{Linton2001,
  author    = {Linton, C.},
  title     = {Handbook of Mathematical Techniques for Wave/Structure Interactions},
  publisher = {Chapman and Hall/CRC Press},
  year      = {2001}
}

@techreport{MacCamyFuchs1954,
  author      = {MacCamy, R. C. and Fuchs, R. A.},
  title       = {Wave forces on piles: a diffraction theory},
  institution = {US Beach Erosion Board},
  year        = {1954}
}

@article{MackayLiangJohanning2021,
  author  = {Mackay, E. and Liang, H. and Johanning, L.},
  title   = {A BEM model for wave forces on structures with thin porous elements},
  journal = {Journal of Fluids and Structures},
  volume  = {102},
  pages   = {103246},
  year    = {2021}
}

@article{MalenicaMolin1995,
  author  = {Malenica, {\v{S}}. and Molin, B.},
  title   = {Third-harmonic wave diffraction by a vertical cylinder},
  journal = {Journal of Fluid Mechanics},
  volume  = {302},
  pages   = {203--229},
  year    = {1995}
}

@article{MaWuEatockTaylor2001a,
  author  = {Ma, Q. and Wu, G. and Eatock-Taylor, R.},
  title   = {Finite element simulation of fully non-linear interaction between vertical cylinders and steep waves. Part 1: methodology and numerical procedure},
  journal = {International Journal for Numerical Methods in Fluids},
  volume  = {36},
  pages   = {265--285},
  year    = {2001}
}

@article{MaWuEatockTaylor2001b,
  author  = {Ma, Q. and Wu, G. and Eatock-Taylor, R.},
  title   = {Finite element simulations of fully non-linear interaction between vertical cylinders and steep waves. Part 2: numerical results and validation},
  journal = {International Journal for Numerical Methods in Fluids},
  volume  = {36},
  pages   = {287--308},
  year    = {2001}
}

@article{MolinKimmounRemyChatjigeorgiou2014,
  author  = {Molin, B. and Kimmoun, O. and Remy, F. and Chatjigeorgiou, I.},
  title   = {Third-order effects in wave--body interaction},
  journal = {European Journal of Mechanics-B/Fluids},
  volume  = {47},
  pages   = {132--144},
  year    = {2014}
}

@inproceedings{MonteserinEngsigKarupEskilsson2018,
  author    = {Monteserin, C. and Engsig-Karup, A. P. and Eskilsson, C.},
  title     = {Nonlinear wave-body interaction using a mixed-Eulerian-Lagrangian spectral element model},
  booktitle = {International Conference on Offshore Mechanics and Arctic Engineering},
  publisher = {American Society of Mechanical Engineers},
  year      = {2018}
}

@article{Morison1950,
  author  = {Morison, J. R. and Johnson, J. W. and Schaaf, S. A.},
  title   = {The force exerted by surface waves on piles},
  journal = {Journal of Petroleum Technology},
  volume  = {2},
  pages   = {149--154},
  year    = {1950}
}

@article{Newman1996,
  author  = {Newman, J. N.},
  title   = {The second-order wave force on a vertical cylinder},
  journal = {Journal of Fluid Mechanics},
  volume  = {320},
  pages   = {417--443},
  year    = {1996}
}

@article{NiedzweckiDuggal1992,
  author  = {Niedzwecki, J. M. and Duggal, A. S.},
  title   = {Wave runup and forces on cylinders in regular and random waves},
  journal = {Journal of Waterway, Port, Coastal, and Ocean Engineering},
  volume  = {118},
  pages   = {615--634},
  year    = {1992}
}

@article{Nielsen2003,
  author  = {Nielsen, F. G.},
  title   = {Comparative study on airgap under floating platforms and run-up along platform columns},
  journal = {Marine Structures},
  volume  = {16},
  pages   = {97--134},
  year    = {2003}
}

@article{NokobYeung2020,
  author  = {Nokob, M. H. and Yeung, R. W.},
  title   = {A fast multipole boundary element method for the three dimensional linear water wave-structure interaction problem with arbitrary bottom topography},
  journal = {Engineering Analysis with Boundary Elements},
  volume  = {117},
  pages   = {232--241},
  year    = {2020}
}

@article{OtsukaIkeda1996,
  author  = {Otsuka, K. and Ikeda, Y.},
  title   = {Estimation of inertia forces on a horizontal circular cylinder in regular and irregular waves at low Keulegan--Carpenter numbers},
  journal = {Applied Ocean Research},
  volume  = {18},
  pages   = {145--156},
  year    = {1996}
}

@article{PanTi2025a,
  author  = {Pan, J. and Ti, Z. and Wang, Y. and Li, Y.},
  title   = {Experimental study on the springing response of a bottom-fixed flexible cylinder subjected to second-order sum-frequency wave forces},
  journal = {Ocean Engineering},
  volume  = {332},
  pages   = {121405},
  year    = {2025}
}

@article{PanTi2025b,
  author  = {Pan, J. and Ti, Z. and Huang, Y. and Li, Y.},
  title   = {Effect of structural flexibility on springing responses of seabed-mounted cylinder under second-order sum-frequency wave forces: Experimental comparisons between rigid and elastic models},
  journal = {Physics of Fluids},
  volume  = {37},
  note    = {in press},
  year    = {2025}
}

@article{PaulsenBredmoseBinghamJacobsen2014,
  author  = {Paulsen, B. T. and Bredmose, H. and Bingham, H. B. and Jacobsen, N. G.},
  title   = {Forcing of a bottom-mounted circular cylinder by steep regular water waves at finite depth},
  journal = {Journal of Fluid Mechanics},
  volume  = {755},
  pages   = {1--34},
  year    = {2014}
}

@article{PaulsenSonnevilleMeulenJacobsen2019,
  author  = {Paulsen, B. T. and {de Sonneville}, B. and {van der Meulen}, M. and Jacobsen, N. G.},
  title   = {Probability of wave slamming and the magnitude of slamming loads on offshore wind turbine foundations},
  journal = {Coastal Engineering},
  volume  = {143},
  pages   = {76--95},
  year    = {2019}
}

@article{RiiseGrueJensenJohannessen2018,
  author  = {Riise, B. H. and Grue, J. and Jensen, A. and Johannessen, T. B.},
  title   = {High frequency resonant response of a monopile in irregular deep water waves},
  journal = {Journal of Fluid Mechanics},
  volume  = {853},
  pages   = {564--586},
  year    = {2018}
}

@article{RoyChakrabortyMandal2017,
  author  = {Roy, R. and Chakraborty, R. and Mandal, B. N.},
  title   = {Propagation of water waves over an asymmetrical rectangular trench},
  journal = {The Quarterly Journal of Mechanics and Applied Mathematics},
  volume  = {70},
  pages   = {49--64},
  year    = {2017}
}

@article{SalisHuLuoRealiManenti2024,
  author  = {Salis, N. and Hu, X. and Luo, M. and Reali, A. and Manenti, S.},
  title   = {3D SPH analysis of focused waves interacting with a floating structure},
  journal = {Applied Ocean Research},
  volume  = {144},
  pages   = {103885},
  year    = {2024}
}

@inproceedings{ShangChenLinWang2026,
  author    = {Shang, D. and Chen, Q. and Lin, X. and Wang, J.},
  title     = {Verifications of a new time-domain model for simulating wave diffraction by vertical cylinder},
  booktitle = {ISOPE International Ocean and Polar Engineering Conference},
  publisher = {ISOPE},
  year      = {2026}
}

@techreport{Shank1970,
  author      = {Shank, G. E.},
  title       = {Assessment of the Morison Equation},
  institution = {Naval Civil Engineering Laboratory},
  number      = {DTIC Report},
  year        = {1970}
}

@article{ShaoFaltinsen2014,
  author  = {Shao, Y.-L. and Faltinsen, O. M.},
  title   = {A numerical study of the second-order wave excitation of ship springing by a higher-order boundary element method},
  journal = {International Journal of Naval Architecture and Ocean Engineering},
  volume  = {6},
  pages   = {1000--1013},
  year    = {2014}
}

@inproceedings{StansbergNielsen2001,
  author    = {Stansberg, C. and Nielsen, F.},
  title     = {Nonlinear wave-structure interactions on floating production systems},
  booktitle = {ISOPE International Ocean and Polar Engineering Conference},
  publisher = {ISOPE},
  year      = {2001}
}

@inproceedings{StansbergBraaten2002,
  author    = {Stansberg, C. and Braaten, H.},
  title     = {Nonlinear wave disturbance around a vertical circular column},
  booktitle = {International Conference on Offshore Mechanics and Arctic Engineering},
  pages     = {857--863},
  year      = {2002}
}

@inproceedings{StansbergBaarholmFokkGudmestadHaver2004,
  author    = {Stansberg, C. T. and Baarholm, R. and Fokk, T. and Gudmestad, O. T. and Haver, S.},
  title     = {Wave amplification and possible deck impact on gravity based structure in $10^{-4}$ probability extreme crest heights},
  booktitle = {International Conference on Offshore Mechanics and Arctic Engineering},
  pages     = {911--920},
  year      = {2004}
}

@article{StansbergKristiansen2005,
  author  = {Stansberg, C. T. and Kristiansen, T.},
  title   = {Non-linear scattering of steep surface waves around vertical columns},
  journal = {Applied Ocean Research},
  volume  = {27},
  pages   = {65--80},
  year    = {2005}
}

@article{SuTengGou2024,
  author  = {Su, G.-f. and Teng, B. and Gou, Y.},
  title   = {Numerical investigation of the fully nonlinear wave force on a vertical cylinder in shallow water using Green-Naghdi theory},
  journal = {Applied Ocean Research},
  volume  = {150},
  pages   = {104117},
  year    = {2024}
}

@article{SuGouZhouTengLi2026,
  author  = {Su, G.-f. and Gou, Y. and Zhou, B. and Teng, B. and Li, Y. P.},
  title   = {Fully nonlinear analysis of wave diffraction by cylinders on a sloping terrain using the Irrotational Green-Naghdi model},
  journal = {Ocean Engineering},
  volume  = {352},
  pages   = {124375},
  year    = {2026}
}

@article{SwanSheikh2015,
  author  = {Swan, C. and Sheikh, R.},
  title   = {The interaction between steep waves and a surface-piercing column},
  journal = {Philosophical Transactions of the Royal Society A: Mathematical, Physical and Engineering Sciences},
  volume  = {373},
  pages   = {20140114},
  year    = {2015}
}

@inproceedings{SwanJohannessenBateman2001,
  author    = {Swan, C. and Johannessen, T. and Bateman, W.},
  title     = {Observations of Extreme 3-D Surface Water Waves},
  booktitle = {Rogue Waves 2000: Proceedings of a Workshop Organized by Ifremer and Held in Brest, France},
  publisher = {Editions Quae},
  pages     = {317},
  year      = {2001}
}

@article{TaylorTangAdcockZang2024,
  author  = {Taylor, P. H. and Tang, T. and Adcock, T. A. and Zang, J.},
  title   = {Transformed-FNV: wave forces on a vertical cylinder—a free-surface formulation},
  journal = {Coastal Engineering},
  volume  = {189},
  pages   = {104454},
  year    = {2024}
}

@article{TengGou2017,
  author  = {Teng, B. and Gou, Y.},
  title   = {BEM for wave interaction with structures and low storage accelerated methods for large scale computation},
  journal = {Journal of Hydrodynamics, Ser. B},
  volume  = {29},
  pages   = {748--762},
  year    = {2017}
}

@article{TengKato1999,
  author  = {Teng, B. and Kato, S.},
  title   = {A method for second-order diffraction potential from an axisymmetric body},
  journal = {Ocean Engineering},
  volume  = {26},
  pages   = {1359--1387},
  year    = {1999}
}

@article{TengKato2002,
  author  = {Teng, B. and Kato, S.},
  title   = {Third order wave force on axisymmetric bodies},
  journal = {Ocean Engineering},
  volume  = {29},
  pages   = {815--843},
  year    = {2002}
}

@article{TongShaoBingham2026,
  author  = {Tong, C. and Shao, Y. and Bingham, H. B.},
  title   = {A weak-scatterer solution to nonlinear wave-body interaction based on a 3-D immersed-boundary adaptive harmonic polynomial cell method},
  journal = {Journal of Hydrodynamics},
  note    = {in press},
  year    = {2026}
}

@article{TrulsenDysthe1997,
  author  = {Trulsen, K. and Dysthe, K. B.},
  title   = {Frequency downshift in three-dimensional wave trains in a deep basin},
  journal = {Journal of Fluid Mechanics},
  volume  = {352},
  pages   = {359--373},
  year    = {1997}
}

@article{VidicPerunovicJensen2005,
  author  = {Vidic-Perunovic, J. and Jensen, J. J.},
  title   = {Non-linear springing excitation due to a bidirectional wave field},
  journal = {Marine Structures},
  volume  = {18},
  pages   = {332--358},
  year    = {2005}
}

@article{Wagner1932,
  author  = {Wagner, H.},
  title   = {Phenomena associated with impacts and sliding on liquid surfaces},
  journal = {Z. Angew. Math. Mech.},
  volume  = {12},
  pages   = {193--215},
  year    = {1932}
}

@article{Wang2025,
  author  = {Wang, J.},
  title   = {An enhanced spectral boundary integral method for modeling highly nonlinear water waves in variable depth},
  journal = {Journal of Computational Physics},
  volume  = {521},
  pages   = {113525},
  year    = {2025}
}

@article{WangMaYan2021,
  author  = {Wang, J. and Ma, Q. and Yan, S.},
  title   = {On extreme waves in directional seas with presence of oblique current},
  journal = {Applied Ocean Research},
  volume  = {112},
  pages   = {102586},
  year    = {2021}
}

@article{WangMaYanLiang2021,
  author  = {Wang, J. and Ma, Q. and Yan, S. and Liang, B.},
  title   = {Modeling crossing random seas by fully non-linear numerical simulations},
  journal = {Frontiers in Physics},
  volume  = {9},
  pages   = {593394},
  year    = {2021}
}

@article{WangMaYangGaoWu2022,
  author  = {Wang, J. and Ma, Q. and Yang, Z. and Gao, J. and Wu, G.},
  title   = {Two types of wave-current interactions and their effects on extreme waves in directional seas},
  journal = {Ocean Engineering},
  volume  = {266},
  pages   = {112637},
  year    = {2022}
}

@article{WestBruecknerJandaMilderMilton1987,
  author  = {West, B. J. and Brueckner, K. A. and Janda, R. S. and Milder, D. M. and Milton, R. L.},
  title   = {A new numerical method for surface hydrodynamics},
  journal = {Journal of Geophysical Research: Oceans},
  volume  = {92},
  pages   = {11803--11824},
  year    = {1987}
}

@article{WienkeOumeraci2005,
  author  = {Wienke, J. and Oumeraci, H.},
  title   = {Breaking wave impact force on a vertical and inclined slender pile—theoretical and large-scale model investigations},
  journal = {Coastal Engineering},
  volume  = {52},
  pages   = {435--462},
  year    = {2005}
}

@article{XiaoLiuWuYue2013,
  author  = {Xiao, W. and Liu, Y. and Wu, G. and Yue, D. K.},
  title   = {Rogue wave occurrence and dynamics by direct simulations of nonlinear wave-field evolution},
  journal = {Journal of Fluid Mechanics},
  volume  = {720},
  pages   = {357--392},
  year    = {2013}
}

@article{XuWang2021,
  author  = {Xu, F. and Wang, Y.},
  title   = {Numerical simulation of ringing responses of a vertical cylinder},
  journal = {Ocean Engineering},
  volume  = {226},
  pages   = {108815},
  year    = {2021}
}

@article{YangStansbyRogersBuldakovStagonasDraycott2023,
  author  = {Yang, Y. and Stansby, P. K. and Rogers, B. D. and Buldakov, E. and Stagonas, D. and Draycott, S.},
  title   = {The loading on a vertical cylinder in steep and breaking waves on sheared currents using smoothed particle hydrodynamics},
  journal = {Physics of Fluids},
  volume  = {35},
  year    = {2023}
}

@article{YangXuTanZhangWangTkalich2026,
  author  = {Yang, Z. and Xu, H. and Tan, W. and Zhang, R. and Wang, J. and Tkalich, P.},
  title   = {Coupled depth-integrated and Navier--Stokes modeling of wave--current--structure interactions},
  journal = {Ocean Engineering},
  volume  = {357},
  pages   = {125425},
  year    = {2026}
}

@article{ZhangelMoctarJiang2026,
  author  = {Zhang, Q. and {el Moctar}, O. and Jiang, C.},
  title   = {Hydroelasticity effects on wave-induced loads for flexible slender components in offshore wind turbines},
  journal = {Marine Structures},
  volume  = {107},
  pages   = {104014},
  year    = {2026}
}

@article{ZhouNingTengBai2013,
  author  = {Zhou, B. and Ning, D. and Teng, B. and Bai, W.},
  title   = {Numerical investigation of wave radiation by a vertical cylinder using a fully nonlinear HOBEM},
  journal = {Ocean Engineering},
  volume  = {70},
  pages   = {1--13},
  year    = {2013}
}

@article{ZhuangWan2020,
  author  = {Zhuang, Y. and Wan, D.},
  title   = {Recent Development of High-Order-Spectral Method Combined with Computational Fluid Dynamics Method for Wave-Structure Interactions},
  journal = {Journal of Harbin Institute of Technology (New Series)},
  volume  = {27},
  year    = {2020}
}

@article{LiuXueYue2001,
  author  = {Liu, Y. and Xue, M. and Yue, D. K.},
  title   = {Computations of fully nonlinear three-dimensional wave--wave and wave--body interactions. Part 2. Nonlinear waves and forces on a body},
  journal = {Journal of Fluid Mechanics},
  volume  = {438},
  pages   = {41--66},
  year    = {2001}
}

@book{MeiStiassnieYue2005,
  author    = {Mei, C. C. and Stiassnie, M. A. and Yue, D. K.-P.},
  title     = {Theory and Applications of Ocean Surface Waves: Part 1: Linear Aspects},
  publisher = {World Scientific},
  year      = {2005}
}

@book{MeiStiassnieYue2018,
  author    = {Mei, C. C. and Stiassnie, M. and Yue, D. K.-P.},
  title     = {Theory and Applications of Ocean Surface Waves},
  edition   = {Third Edition (In 2 Volumes)},
  publisher = {World Scientific},
  doi       = {10.1142/10212},
  year      = {2018}
}

@article{XueXuLiuYue2001,
  author  = {Xue, M. and X{\"u}, H. and Liu, Y. and Yue, D. K.},
  title   = {Computations of fully nonlinear three-dimensional wave--wave and wave--body interactions. Part 1. Dynamics of steep three-dimensional waves},
  journal = {Journal of Fluid Mechanics},
  volume  = {438},
  pages   = {11--39},
  year    = {2001}
}

@article{ClamondGrue2001,
  author  = {Clamond, D. and Grue, J.},
  title   = {A fast method for fully nonlinear water-wave computations},
  journal = {Journal of Fluid Mechanics},
  volume  = {447},
  year    = {2001}
}

@article{FructusEtAl2005,
  author  = {Fructus, D. and Clamond, D. and Grue, J. and Kristiansen, {\O}.},
  title   = {An efficient model for three-dimensional surface wave simulations: Part I: Free space problems},
  journal = {Journal of Computational Physics},
  volume  = {205},
  year    = {2005}
}

@article{GrilliSkourupSvendsen1989,
  title   = {An efficient boundary element method for nonlinear water waves},
  author  = {Grilli, St{\'e}phan T. and Skourup, J. and Svendsen, Ib A.},
  journal = {Engineering Analysis with Boundary Elements},
  volume  = {6},
  number  = {2},
  pages   = {97--107},
  year    = {1989},
  doi     = {10.1016/0955-7997(89)90005-2}
}

@article{SungChoi2010,
  title   = {Implicit formulation with the boundary element method for nonlinear radiation of water waves},
  author  = {Sung, Hong Gun and Choi, Hang Shoon},
  journal = {Engineering Analysis with Boundary Elements},
  volume  = {34},
  number  = {5},
  pages   = {511--529},
  year    = {2010},
  doi     = {10.1016/j.enganabound.2009.11.005}
}

@article{GuerberBenoitGrilliBuvat2012,
  title   = {A fully nonlinear implicit model for wave interactions with submerged structures in forced or free motion},
  author  = {Guerber, Etienne and Benoit, Michel and Grilli, St{\'e}phan T. and Buvat, Cl{\'e}ment},
  journal = {Engineering Analysis with Boundary Elements},
  volume  = {36},
  number  = {7},
  pages   = {1151--1163},
  year    = {2012},
  doi     = {10.1016/j.enganabound.2012.02.005}
}

@article{ZhouWuMeng2016,
  title   = {Interactions of fully nonlinear solitary wave with a freely floating vertical cylinder},
  author  = {Zhou, Binzhen and Wu, G. X. and Meng, Q. C.},
  journal = {Engineering Analysis with Boundary Elements},
  volume  = {69},
  pages   = {119--131},
  year    = {2016},
  doi     = {10.1016/j.enganabound.2016.05.004}
}

@article{TsaoKees2023,
  title   = {An Arbitrary Lagrangian--Eulerian Regularized Boundary Integral Method for Nonlinear Free-Surface Flows over Complex Topography and Wave-Structure Interaction},
  author  = {Tsao, Wen-Huai and Kees, Christopher E.},
  journal = {Engineering Analysis with Boundary Elements},
  volume  = {157},
  pages   = {326--341},
  year    = {2023},
  doi     = {10.1016/j.enganabound.2023.09.019}
}

\end{document}